\documentstyle[onecolumn]{mn}
\newcommand{\mincir}{\raise -2.truept\hbox{\rlap{\hbox{$\sim$}}\raise5.truept
\hbox{$<$}\ }}
\newcommand{\magcir}{\raise -2.truept\hbox{\rlap{\hbox{$\sim$}}\raise5.truept
\hbox{$>$}\ }}
\newcommand{\minmag}{\raise-2.truept\hbox{\rlap{\hbox{$<$}}\raise 6.truept\hbox
{$>$}\ }}
\newcommand{\be}{\begin{equation}}
\newcommand{\ee}{\end{equation}}
\newcommand{\ba}{\begin{eqnarray}}
\newcommand{\ea}{\end{eqnarray}}
\newcommand{\brr}{\begin{array}}
\newcommand{\err}{\end{array}}
\newcommand{\bc}{\begin{center}}
\newcommand{\ec}{\end{center}}

\input{psfig.sty}
\title{Modelling galaxy clustering at high redshift}
\author[Moscardini et al.]
{Lauro Moscardini$^{1}$,
Peter Coles$^{2}$,
Francesco Lucchin$^{1}$ and
Sabino Matarrese$^{3}$\\
$^1$Dipartimento di Astronomia, Universit\`a di Padova,
vicolo dell'Osservatorio 5, I--35122 Padova, Italy\\
$^2$Astronomy Unit, School of Mathematical Sciences, 
Queen Mary \& Westfield College, Mile End Road, 
London E1 4NS\\
$^3$Dipartimento di Fisica G. Galilei,
Universit\`{a} di Padova, via Marzolo 8, I--35131 Padova, Italy}
\begin{document}

\maketitle

\begin{abstract}
We discuss the theoretical interpretation of observational data concerning the
clustering of galaxies at high redshifts. Building on the  theoretical
machinery developed by Matarrese et al. (1997), we make detailed quantitative
predictions of galaxy clustering statistics for a variety of cosmological
models, taking into account differences in spatial geometry and initial
fluctuation spectra and exploring the role of bias as a complicating factor in
these calculations. We demonstrate that the usual description of evolution (in
terms of the parameters $\epsilon$ and $r_0$) is not useful for realistic
galaxy clustering models. We compare the detailed predictions of the variation
of correlation functions with redshift against current observational data to
constrain available models of structure formation. Theories that fit the
present-day abundance of rich clusters are generally compatible with the
observed redshift evolution of  galaxy clustering if galaxies are no more than
slightly biased at $z\sim 1$. We also discuss the interpretation of a
concentration of Lyman-break galaxies found by Steidel et al. (1998), coming to
the conclusion that such concentrations are not unexpected in `standard' models
of structure formation. 
\end{abstract}

\begin{keywords}
cosmology: theory -- cosmology: observations -- large--scale structure
of Universe -- galaxies: formation -- galaxies: evolution -- galaxies: haloes
\end{keywords}

\section{Introduction}

The direct observation of statistically complete samples of galaxies at high
redshift has only recently become technologically feasible, but it promises to
yield important information about the origin and evolution of cosmic
structures. In particular, the possibility to observe ``normal" galaxies (or
their progenitors), rather than just quasars or other ultraluminous sources,
may lead to an understanding of how these objects relate to the distribution of
the dark matter which is assumed, in most theories, to dominate the density of
the Universe. This, in turn, should allow observations of galaxy clustering to
be used to gain insights about fundamental aspects of cosmological models, as
well as learning about the way galaxies themselves evolve. But in this field,
theory currently lags considerably behind observations, with the result that
experimental data are sometimes placed in a naive, or perhaps simply incorrect,
theoretical context. 

In Matarrese et al. (1997; hereafter Paper I), we discussed high-redshift
clustering phenomena from a theoretical perspective. In particular, we
developed a general formalism which one can use to make detailed predictions of
statistical measures of clustering and which also makes explicit the main
sources of theoretical uncertainty in these predictions. This formalism allows
one to make a realistic assessment of how models of structure formation fare in
the face of results from particular observational programmes. This formalism is
considerably more complicated than the usual simple scaling ansatz which forms
the framework within which most observational results have previously been
interpreted. 

It is the main purpose of this paper to deploy the techniques of Paper I in a
systematic comparison of currently popular structure formation models with the
available observational data. The models are all based on the cold dark matter
(CDM) model, but vary in the amount of dark matter, the initial perturbation
spectrum, the background cosmology and in the presence or absence of a
cosmological constant. These variations have been introduced in an attempt to
reconcile the basic idea of CDM with cosmological observations which rule out
the simplest version of the model; see, e.g. Coles (1996). This detailed
comparison requires us to specify the assumptions entering our calculations
very carefully, so we also take the opportunity to refine the general arguments
we made in Paper I. 

The layout of the paper is as follows. In Section 2, we briefly recap the main
elements of the approach outlined in Paper I and explain the extension to, for
example, the case of an open universe or models with cosmological constant. We
also describe the six models of structure formation we use to compare with the
data.  We explore the issue of bias in Section 3, drawing on some themes that
were introduced in Paper I, but focussing on the specific case of galaxy
formation in hierarchical clustering models. Section 4 contains a critique of
some simple theoretical arguments often presented in the literature. The
detailed comparison of clustering observations with our theoretical predictions
is made in Section 5, where the relevant data are also described. We also
comment on the theoretical interpretation of a concentration of galaxies found
at redshift $\sim 3$ by Steidel et al. (1998). We present our main conclusions
in Section 6. 

\section{Modelling the evolution of clustering} 

\subsection{Preliminaries} 
In order to model the clustering of high--redshift galaxies, and to compare
observational data with the predictions of different cosmological models,  one
must confront a number of different issues. Firstly, there is the necessity to
have a reliable way to follow the redshift evolution of matter correlations.
Secondly, one has to model the relationship between fluctuations in the mass
and fluctuations in observable galaxies, and to consider the possible time
evolution of this relationship. In the usual language, this means devising a
model for the {\em bias}. Another ingredient is the calculation of effects
introduced by the observing process, such as the role of the selection
function, and the possible effect of redshift-space distortions on measures of
clustering. Finally, because of the limited number of galaxies in available
samples and the need to obtain a reasonable statistical signal-to-noise ratio,
observational results are usually presented for galaxies spanning a  range of
redshifts. This last effect means, for example, that the observed correlation
function of the sample  involves a convolution of the real correlation function
with the redshift distribution of the objects contained in the sample. 

We should also point out that gravitational lensing effects along the
observer's past light cone also introduce a bias into the observed statistics
(e.g. Villumsen 1996; Moessner, Jain \& Villumsen 1998). The presence of a
correlated shear results in increased apparent clustering over and above that
produced by the intrinsic galaxy correlations. Likewise the distortion
introduced by inferring galaxy positions from redshifts also acts to increase
clustering statistics with respect to the real-space versions, except for
projected correlations which do not suffer from this effect. Since these
effects both act in the same direction, calculations made without taking them
into account are not exactly comparable with observations: more realistic
predictions, however, would always be higher than those we present here. 

In Paper I we developed a formalism that takes into account all these
requirements, and most of this section is devoted to a brief summary of the
machinery that was constructed there. Essentially, Paper I showed that the
observed correlation function $\xi_{\rm obs}$ in a given redshift interval
${\cal Z}$ is an appropriate weighted average of the mass autocorrelation
function $\xi$ with the mean number of objects ${\cal N}$ and effective bias
factor $b_{\rm eff}$, defined below in equation (\ref{eq:b_eff}), in that
range: 
\be
\xi_{\rm obs}(r) = N^{-2}
\int_{\cal Z} d z_1 dz_2
~{\cal N}(z_1) ~{\cal N}(z_2) ~b_{\rm eff}(z_1) ~b_{\rm eff}(z_2)
~\xi(r,\bar z) \;,
\label{eq:xifund}
\ee
where $N \equiv \int_{\cal Z} d z' {\cal N}(z')$ and $\bar z$ is an
intermediate redshift between $z_1$ and $z_2$. Porciani (1997), by computing
the evolution of the two-point correlation function in the Zel'dovich (1970)
approximation, showed that adopting the relation $D_+(\bar{z})=\sqrt{D_+(z_1)
D_+(z_2)}$, where $D_+$ is the growth law for linear density fluctuations (see
below, Section 2.4), ensures that predictions will be accurate to an error
smaller than 1 per cent. In the following we will assume this expression to
define $\bar z$. 

\subsection{Bias and all that} 
The factor of $b_{\rm eff}$ which appears in equation (\ref{eq:xifund}) is a
consequence of our lack of understanding of the details of the galaxy formation
process and the consequently uncertain relationship between fluctuations in
matter density $\delta_{\rm m}$ and galaxy number-density $\delta_{\rm n}$. It
is conventional to parametrise one's ignorance in this arena by introducing a
single linear {\em bias parameter} such that a relationship of the form
$\delta_{\rm n}=b\, \delta_{\rm m}$ is assumed. In this work we shall
generalise this idea so that we assume that objects with given intrinsic
properties (such as mass $M$) and at different redshifts $z$ can have different
bias parameters, which we call $b(M,z)$. For each set of objects, however, the
bias is assumed still to be linear; it is also {\em local}, in the sense that
the propensity of galaxies to form at a given spatial location ${\bf x}$
depends only on the matter density at that point: 
\be
\delta_{\rm n}({\bf x}; M,z) \simeq b(M,z) \delta_{\rm m}({\bf x},z);
\label{eq:bialoc}
\ee
so that no environmental or co-operative effects in galaxy formation are
permitted. If we  assume such a bias between the galaxy and mass fluctuations,
the {\em effective} bias factor $b_{\rm eff}(z)$ which appears in equation
(\ref{eq:xifund}) can be expressed as a suitable average of the
``monochromatic'' bias $b(M,z)$ (i.e. the bias factor of each single object): 
\be
b_{\rm eff}(z) \equiv {\cal N}(z)^{-1} \int_{\cal M} d\ln M' ~b(M',z)
~{\cal N}(z,M')\; .
\label{eq:b_eff}
\ee
Here the variable $M$ (and its range ${\cal M}$) does not necessarily simply
represent the object's mass but rather it stands for any generic intrinsic
properties of the object (mass, luminosity, etc.) on which the selection of the
object into an observational sample might depend. 

Because it plays such an important role in this formalism, we have devoted the
whole of Section 3 to a more detailed discussion of the possible form of a bias
and its redshift evolution. 

\subsection{Clustering Statistics} 
Owing to the relatively small size  of the datasets available at the present
time, clustering properties of high-redshift galaxies are generally studied in
terms of the angular [$\omega_{\rm obs}(\vartheta)$] or of the projected
real-space [$w_{\rm obs}(r_p)$] correlation functions. Adopting the {\em
small--angle} approximation, in Paper I we obtained for $\omega_{\rm obs}$: 
\be
\omega_{\rm obs}(\vartheta) = N^{-2}
\int_{\cal Z} d z ~G(z) ~{\cal N}^2(z) ~b^2_{\rm eff}(z) \int_{-\infty}^\infty
d u ~\xi[r(u,\vartheta,z) , z] \;,
\ee
where $r(u,\vartheta,z) \equiv a_0 \sqrt{u^2 + x^2(z) \vartheta^2}$, with
$x(z)$ given by
\be
x(z)={c\over{H_0 a_0 \sqrt{|\kappa |}}} ~{\cal S}\left(\sqrt{|{\kappa}|}
\int_0^z \left[ \left( 1+z' \right)^2 \left(1+\Omega_{0m} z'\right) -
z'\left(2+z'\right) \Omega_{0\Lambda}\right]^{-1/2} dz'\right)
\label{eq:x_z}
\ee
and
\be
G(z) \equiv \biggl({d x \over d z}\biggr)^{-1} \;.
\ee
In equation (\ref{eq:x_z}) and hereafter, we use $\Omega_{0m}$ and
$\Omega_{0\Lambda}$ to represent the contribution at the present time to a
critical energy density from matter and vacuum energy respectively. When we
require these quantities at an arbitrary time we use $\Omega_m$ and
$\Omega_\Lambda$; since they evolve with epoch, $\Omega_m$ and $\Omega_\Lambda$
are implicit functions of $z$; although the cosmological constant $\Lambda$ is
constant in redshift, $\Omega_\Lambda=\Lambda/3H^{2}$, which varies through the
Hubble constant $H(z)$. We write the total density parameter $\Omega_m+
\Omega_\Lambda\equiv\Omega_t$; consequently $\Omega_{0m}+\Omega_{0\Lambda}=
\Omega_{0t}$. 

Note also that, while in the Einstein--de Sitter case $a_0$ is an arbitrary
length--scale which can be set to unity, in the non--flat case it is given by 
\be
a_0 = \frac{c}{H_0} \vert 1 - \Omega_{0t} \vert^{-1/2} \;.
\ee
In equation (\ref{eq:x_z}), if $\Omega_{0t}< 1$, ${\cal S}(x)\equiv\sinh(x)$
and $\kappa=1-\Omega_{0t}$; if $\Omega_{0t}> 1$, ${\cal S}(x)\equiv\sin(x)$ and
$\kappa=1-\Omega_{0t}$; if $\Omega_{0t}=1$, ${\cal S}(x)\equiv x$ and
$\kappa=1$. In the case of a vanishing cosmological constant, the previous
expression can be solved analytically and can be written as 
\be
x(z) = {2c \over a_0 H_0}
{\Omega_{0m} z + (\Omega_{0m}-2) [ - 1 + (\Omega_{0m} z + 1)^{1/2}]
\over \Omega_{0m}^2 (1+z)}\;.
\ee
The {\em projected} real--space correlation function $w_{\rm obs} $ can be
directly obtained by $\xi_{\rm obs}(r)$ as 
\be
w_{\rm obs}(r_p)
= 2 \int_0^\infty d y ~\xi_{\rm obs}(\sqrt{r_p^2+y^2}) =
2 \int_{r_p}^\infty d r ~r ~(r^2 - r_p^2)^{-1/2}
~\xi_{\rm obs}(r) \;,
\ee
where $r_p$ is the component of the pair separation perpendicular to the line
of sight. 

\subsection{Evolution of the mass autocorrelation function} 
The non-linear growth of the density fluctuations modifies the shape of the
power spectrum $P(k)$ as well as its amplitude. In the linear regime, which
holds at large scales and/or early times, the solution is given by the relation
\be
P(k,z)= D^2_+(z) P(k,z=0)\ ,
\ee
where $D_+(z)$ is the growing mode of linear perturbations normalised to unity
at $z=0$, so that the spectrum grows with time without any change in shape. By
contrast, in the strongly non-linear regime, some theoretical arguments and
numerical simulation results suggest the existence of the so-called stable
clustering regime  wherein the matter correlations exhibit a particular form of
self-similar behaviour (Peebles 1980; Jain, Mo \& White 1995; Jain 1997),
although whether the stable clustering description holds in detail is still
open to some doubt (e.g. Padmanabhan et al. 1996; Munshi et al. 1997). We shall
return to this issue later, in Section 4 of this paper. In any case, the
non-linear behaviour of matter correlations on small scales results in a
distortion of the shape of the matter power spectrum from its initial form. 

The clustering behaviour which is most relevant to the study of objects at high
redshift is actually in between these two extremes. This intermediate regime is
generally studied by fitting results from numerical simulations with a
semi-empirical universal function, obtained by following the simple {\em
ansatz} originally introduced by Hamilton et al. (1991). Recognising  that
gravitational collapse changes the effective length scale of a density
perturbation, Hamilton et al. (1991) suggested that the initial (linear) scale
$r_0$ of a density perturbation should be related to the final (non-linear)
scale $r$ of the same perturbation after collapse by: 
\be
r_0=[1+\bar \xi(r,z)]^{1/3} r \;,
\label{eq:r_0}
\ee
where
\be
\bar \xi(r,z) \equiv {3 \over r^3} \int_0^r y^2 \xi(y,z) dy \;
\ee
is the integrated correlation function. The Hamilton et al. idea is that there
is a universal function $F$ acting on $\bar \xi$, once the change in
appropriate length scale is taken into account: 
\be
\bar \xi(r,z) = F[\bar \xi_{\rm lin}(r_0,z)]\; .
\label{eq:xirz}
\ee
Recently this formalism has been significantly refined and generalised. Peacock
\& Dodds (1994) considered the application of this method to the power spectra
rather than the correlation function, and also to models with arbitrary
background density (i.e. with $\Omega_{0t}\ne 1$ or $\Omega_{0\Lambda}\ne 0$).
Jain, Mo \& White (1995) introduced the dependence of the universal function
$F$ of the primordial spectral index $n$. Finally Peacock \& Dodds (1996,
hereafter PD96), by using N-body simulations with high spatial resolution,
obtained accurate fits for $F$ both in low-density universes and universes with
non-vanishing cosmological constant. The accuracy of this form of the fitting
function has been recently confirmed using  numerical experiments by Jenkins et
al. (1997). 

In Paper I we followed the clustering evolution by using the fitting function
obtained by Jain, Mo \& White (1995). Here, because it is our intention  to
consider models with $\Omega_{0t} \ne 1$ or 
$\Omega_{0\Lambda} \ne 0$, we have decided instead to use the
form of the method presented by PD96 which deals with the (dimensionless) power
spectrum $\Delta^2$: 
\be
\Delta^2(k)\equiv {{d\sigma^2}\over{d \ln k}}={1\over {2\pi^2}} k^3 P(k)\ ,
\ee
which is related to the two-point correlation function by
\be
\xi(r) = \int \Delta^2(k) {{\sin kr}\over{kr}} {{dk}\over{k}}\ ;
\ee
$\sigma^2$ is the variance of the density fluctuation field. In this case, the
corresponding expressions for equations (\ref{eq:xirz}) and (\ref{eq:r_0}) are:
\be
\Delta^2(k,z) = {\cal F}[\Delta^2_{\rm lin}(k_0,z)]\;, \ \
\ \ \ \
k_0=[1+\Delta^2(k,z)]^{-1/3} k \;,
\ee
where $k_0$ and $k$ are the linear and non-linear wavenumbers, respectively. We
adopt the universal fitting function ${\cal F}$ given by PD96, which depends on
$g$, a suppression factor that measures the rate of clustering growth in a
particular cosmology relative to that which pertains in a flat matter-dominated
Universe. The quantity $g$ therefore contains a dependence on the background
cosmology. Carroll, Press \& Turner (1992) found an approximate (but almost
exact) expression for $g$: 
\be
g(\Omega_m, \Omega_\Lambda)={5\over 2} \Omega_m [\Omega_m^{4/7}-\Omega_\Lambda
+
(1+\Omega_m/2)(1+\Omega_\Lambda/70)]^{-1}\ ;
\ee
Since $\Omega_m$ and $\Omega_\Lambda$ both depend upon redshift $z$ (e.g.
Section 2.3), and it is the dependence of $g$ on $z$ in which we are
interested, we can use $g(z)$ to encode this behaviour for any particular
cosmological model. The formula given for ${\cal F}$ by PD96 was originally for
$z=0$, but it applies at any cosmic epoch $z$ by replacing $g$ by $g(z)$. The
quantity $x$ must also be interpreted as the linear power spectrum at the epoch
$z$, i.e. 
\be
x=\Delta^{2}_{\rm lin} (k_0, z)= \Delta^{2}_{\rm lin} (k_0, z=0)
(1+z)^{-2} [g(z)/g(0)]^{2}\, ,
\ee
rather than at $z=0$ which is assumed in the original PD96 application. 

The form of ${\cal F}(x)$ given by PD96 assumes a power-law initial spectrum
described by an index $n$. For models which are not described by pure power-law
spectra, such as the variations on the cold dark matter model that we shall
discuss in this paper, one can use the same formulae, but replacing $n$ by an
effective index $n_{\rm eff}$, defined by 
\be
n_{\rm eff}(k_0,z) = \frac{d \ln P_{\rm lin}(k,z)}
{d \ln k} \big|_{k=k_0(z)/2}\ .
\ee
PD96 claim that this prescription is able to reproduce the non-linear evolution
with a precision of about 7 per cent, which is perfectly adequate for the
application we have in mind for this paper. 

Some care has to be taken if one seeks  to apply the PD96 method to
cosmological models where $\Delta^2$ displays a significant peak at some
particular wavenumber. Such models are not really consistent with the
assumption of hierarchical clustering in the first place. For example, tilted
cold dark matter models have $n_{\rm eff}<-3$ for large $k$, which means that
they have a very sharp spectral feature and are susceptible to this difficulty
(see e.g. Vittorio, Matarrese \& Lucchin 1988). Of course, the method can still
be used to construct the power spectra on large scales and it is possible to
use safely its predictions down to the linear wavenumber where the spectrum
peaks. For many variants of tilted cold dark matter models (as those described
in the following subsection), $\Delta^2$ is sufficiently large there that the
non-linear power will be very large. One would not want to make predictions
based solely on dark matter clustering to any smaller scales, so the fact that
$n_{\rm eff} < -3$ on the smallest linear scales is not a problem in the
context of this work. 

\subsection{The cosmological models} 
One of the limitations of Paper I was that it restricted attention to a simple
phenomenological model of the initial power spectrum and to a flat spatial
geometry. It is one of the aims of this paper to extend the treatment to study
a wider range of initial conditions and relevant changes in the global
cosmological parameters (including spatial curvature). This is a particularly
interesting task at the present time because it is well known that the
so--called standard cold dark matter (SCDM) model --- which assumes a flat
universe with $\Omega_0=1$ and $\Omega_\Lambda=0$, a spectral index $n=1$, a
Hubble constant (in units of 100 km s$^{-1}$ Mpc$^{-1}$) $h=0.5$ and a baryon
fraction $\Omega_b=0.0125 h^{-2}$, as predicted by the standard theory of the
big bang nucleosynthesis --- is not able to reproduce the clustering properties
of the galaxy and cluster distribution and the cluster abundances, when
normalized to the COBE data. As a consequence, a number of variants on this
basic scenario have been suggested  which might remedy its shortcomings. In
this paper we consider different cosmological models which might be viable
alternatives to the SCDM model; they all have a similar basic shape of power
spectrum to SCDM but are engineered to have a smaller amount of small-scale
power, which is the main problem with SCDM itself. In a general way, the
initial (linear regime) power spectrum for all these models can be represented
by 
\be
P_{\rm lin}(k,0)=P_0 k^n T^2(k)\ ,
\ee
where we use the fitting formula of the CDM transfer function as given by
Bardeen et al. (1986):
\be
{T(k)}={{\ln(1+2.34q)}\over {2.34q}} \left[ 1+3.89q+(16.1q)^2+(5.46q)^3+
(6.71q)^4\right]^{-1/4}\ .
\ee
In the previous equation $q\equiv (k/h \ {\rm Mpc}^{-1})/\Gamma$. The shape
parameter $\Gamma$ is related to the matter density parameter $\Omega_{0m}$ and
to the baryonic fraction $\Omega_{0b}$ by the relation $\Gamma=\Omega_{0m} h
\exp[-\Omega_{0b}-\sqrt{h/0.5}\ \Omega_{0b}/\Omega_{0m})]$ (Sugiyama 1995). 

To fix the amplitude of the power spectrum, we either attempt to fit the
present-day cluster abundance or the level of fluctuations observed by COBE.
For the latter, we parametrise the normalisation of the 4--year COBE data in
terms of  $\sigma_8$, the r.m.s. fluctuation amplitude inside a sphere of $8
h^{-1}$ Mpc, using the results of Bunn \& White (1997), who used a
Karhunen-Lo\`{e}ve expansion to produce an unbiased estimate of the
normalization, with a statistical uncertainty reduced to 7 per cent. 

We will consider the following specific models, the main parameters of which
are described in Table 1: 

\begin{itemize}
\item
the SCDM model, as reference model, with a normalization consistent with the
COBE data;
\item
a different version of the SCDM model, hereafter called SCDM$_{CL}$, with a
reduced normalization corresponding to $\sigma_8=0.52$ which produces a cluster
abundance in better agreement with the observational data (Eke, Cole \& Frenk
1996; see also Viana \& Liddle 1996); 
\item
a tilted model (hereafter TCDM; see e.g. Lucchin \& Matarrese 1985; Vittorio,
Matarrese \& Lucchin 1988) with $n=0.8$ and high baryonic content
($\Omega_{0b}=0.1$; see White et al. 1996; Gheller, Pantano \& Moscardini
1998); 
\item
a different version of the previous model, hereafter TCDM$_{GW}$, with a
reduced normalization of the scalar perturbations ($S$) that takes into account
the possible production of gravitational waves (tensor perturbations $T$) to
the COBE fluctuations [we adopt the ratio $T/S=7(1-n)$ for the ratio of tensor
to scalar contribution to the quadrupole, as predicted by some inflationary
theories (e.g. Lucchin, Matarrese \& Mollerach 1992; Lidsey \& Coles 1992)]; 
\item
a open CDM model, with $\Omega_{0t}=\Omega_{0m}=0.4$, COBE--normalized
(hereafter OCDM); 
\item
a low--density CDM model ($\Omega_{0m}=0.4$), with flatness provided by the
cosmological term, i.e. $\Omega_{0t}=1$ and $\Omega_{0\Lambda}=1-\Omega_{0m}$,
COBE--normalized (hereafter $\Lambda$CDM). 
\end{itemize}

\begin{table}
\centering
\caption[]{The parameters of the cosmological models. Column 2: the present
matter density
parameter $\Omega_{0m}$; Column 3: the present
cosmological constant contribution to the
density $\Omega_{0\Lambda}$; Column 4: the primordial spectral index
$n$; Column 5: the Hubble parameter $h$; Column 6: the present baryon density
$\Omega_{b}$; Column 7: the shape parameter $\Gamma$; Column 8: the spectrum
normalization $\sigma_8$; Column 9: the non-linear value of the r.m.s.
fluctuation amplitude inside a sphere of $8 h^{-1}$ Mpc $\sigma_8^{nl}$.}
\tabcolsep 4pt
\begin{tabular}{lcccccccc} \\ \\ \hline \hline
Model & $\Omega_{0m}$ & $\Omega_{0\Lambda}$ & $n$ & $h$ & $\Omega_{0b}$ &
$\Gamma$ & $\sigma_8$ & $\sigma_8^{nl}$ \\ \hline
SCDM         & 1.0 & 0.0 & 1.0 & 0.50 & 0.050 & 0.45 & 1.22 & 1.16 \\
SCDM$_{CL}$  & 1.0 & 0.0 & 1.0 & 0.50 & 0.050 & 0.45 & 0.52 & 0.51 \\
TCDM         & 1.0 & 0.0 & 0.8 & 0.50 & 0.100 & 0.41 & 0.72 & 0.72 \\
TCDM$_{GW}$  & 1.0 & 0.0 & 0.8 & 0.50 & 0.100 & 0.41 & 0.51 & 0.51 \\
OCDM         & 0.4 & 0.0 & 1.0 & 0.65 & 0.036 & 0.23 & 0.64 & 0.66 \\
$\Lambda$CDM & 0.4 & 0.6 & 1.0 & 0.65 & 0.036 & 0.23 & 1.07 & 1.13 \\
\hline
\end{tabular}
\label{t:models}
\end{table}

\section{Modelling the bias of galaxies}
Though the number of ingredients is large in the models introduced above, the
theoretical understanding of how clustering of matter grows via gravitational
instability in the expanding Universe is quite well developed. As a
consequence, it is relatively straightforward to compute the autocovariance
function of matter fluctuations as a function of redshift in these scenarios.
As we mentioned in Section 2.2, however, this does not lead us directly to a
prediction of galaxy correlation properties because we still do not fully
understand the details of the relationship between the whereabouts of the
galaxies and the whereabouts of the mass. In principle, this relationship could
be highly complicated, non-linear and  environment-dependent. If this turns out
to be the case then it is going to be very difficult indeed ever to unravel
galaxy clustering observations to obtain information about the evolution of
matter fluctuations and the cosmological parameters on which they depend. In
this spirit, we are motivated to assume the relatively simple form of local
bias represented by equation (\ref{eq:bialoc}), though we do admit at the
outset that things could be much more complex than this. 

Having settled on equation (\ref{eq:bialoc}), our task is now to determine the
behaviour of the function $b(M,z)$ for a given theoretical picture. In Paper I,
we introduced four different general ideas of how different classes of cosmic
objects might be related to the mass distribution and parametrised them in
terms of the simplified biasing model we adopted. In this paper, we shall stick
to the same four basic models, but adapt them to the specific situation of
galaxy clustering. 

The simplest biasing model one can imagine, and which is regarded by many as
the most realistic, is described by $b(M,z)=1$. This is called the {\em
unbiased} model, though one has to be a little careful in using this
terminology and motivation for it, particularly at high redshifts, is actually
quite limited. In fact, it is not a trivial question to ask what is the formal
definition of an unbiased population of objects, since one is attempting to
relate two different types of mathematical field: a point set and a continuous
density field. One useful definition is that the population of objects
constitutes a random (Poisson) sampling of the matter distribution, such as is
the case if galaxies are selected by their luminosity which, in turn, is drawn
from a universal (position-independent) luminosity function. But as galaxy
formation occurs, stellar populations and luminosities evolve and galaxies
undergo merging and tidal disruption, it is difficult to see how $b(M,z)$ can
be equal to unity for all properties $M$ and redshifts $z$. In particular, at
sufficiently high $z$ one reaches the point where the first galaxy to form in
the observable universe produced its stars. It clearly makes no sense to
describe this object by an unbiased model in the sense we have used it here,
even if one does not invoke density-dependent luminosity functions or other
environmental effects. Moreover, galaxies in general must represent some form
of subset of the population of {\em collapsed} objects and, as we discuss later
on, these objects are generally biased with respect to the underlying continuum
mass distribution. So even though it is very simple and, at least at first
sight, self-consistent, we do not believe this model to be well motivated in
the context of this paper so we use it here as a reference for the more complex
models which follow. 

An alternative picture of biasing can be constructed by imagining that galaxy
formation occurs, for a given class of galaxies, at a relatively well-defined
redshift $z_f$. (One can either assume that there is a single typical formation
redshift for a certain class of galaxies or that there is some spread in it.)
If this is the case, one can further imagine that galaxies which are born at a
given epoch $z_f$ might well be imprinted with a particular value of
$b(M,z_f)$, in the spirit of equation (\ref{eq:bialoc}), as long as the
formation event is relatively local. If galaxies are biased by birth in this
way, then they will not continue with the same biasing factor for all time, but
will tend to be dragged around by the surrounding density fluctuations, which
are perhaps populated by objects with a smaller bias parameter. In this case,
the evolution of the bias factor can be obtained from (Dekel 1986; Dekel \&
Rees 1987; Nusser \& Davis 1994; Fry 1996) 
\be
b(z)= 1+ (b_f-1) { D_+(z_f)  \over D_+(z) } \;, \ \ \ \ \ \ \ \ \ z<z_f \;,
\label{eq:b_cons}
\ee
where $b_f$ is the bias at the formation redshift $z_f$; we have suppressed the
dependence on $M$ here for brevity.  This was called {\em galaxy-conserving
model} in Paper I. Again, it is difficult to motivate this model in detail
because it is difficult to believe that all galaxies survive intact from their
birth to the present epoch, but it at least gives a plausible indication of the
sense in which one expects $b$ to evolve if the timescale for galaxy formation
is relatively short and the timescale under which merging or disruption occurs
is relatively long. 

In most fashionable models of structure formation, however, the growth of
large-scale features is driven by the hierarchical merging of sub-units. In
these theories, one would not expect the survival of galaxies in their pristine
initial state as anticipated in the galaxy-conserving model. Since the
development of the clustering hierarchy is driven by gravity, the first things
one has to understand are the properties of galactic haloes rather than the
galaxies residing in them. One begins by calculating the bias parameter
$b(M,z)$ for haloes of mass $M$ and `formation redshift' $z_f$ at redshift
$z\leq z_f$ in a given cosmological model. The result is 
\be
b(M,z\vert z_f) = 1 + {D_+(z_f) \over \delta_c D_+(z)}
\biggl( {\delta_c^2 \over \sigma_M^2 D_+(z_f)^2 } - 1\biggr) \;,
\label{eq:b_mono}
\ee
where $\sigma^2_M$ is the linear mass-variance averaged over the scale $M$
extrapolated to the present time ($z=0$) and $\delta_c$ is the critical linear
overdensity for spherical collapse [$\delta_c={\rm const}=1.686$ in the
Einstein--de Sitter case, while it depends slightly on $z$ for more general
cosmologies (Lilje 1992)]. The above expression for the bias parameter was
originally calculated by Mo \& White (1996), although for simplicity they only
gave results for $z=0$. The general non-linear relation between the halo and
the mass density contrast has been recently obtained by Catelan et al. (1998),
by solving the continuity equation for dark matter haloes. Bagla (1997b) has
discussed the clustering of haloes using numerical experiments. See also Ogawa,
Roukema \& Yamashita (1997) for a related discussion. 

At this point one has to make some assumptions on how the galaxy is connected
to the hosting halo and on what happens when the halo merges with other haloes.
This point has been discussed at some length in the literature (e.g. Kauffmann,
Nusser \& Steinmetz 1997; Roukema et al. 1997) and many issues still remain
unresolved: it is one of the most complicated aspects of galaxy formation. In
order to make progress we shall simply assume in what follows that, however
star formation and stellar evolution proceeds in a halo once it has formed, the
properties of the resulting galaxy are in a one-to-one relationship with the
parent halo mass $M$. Using this assumption it now becomes clear that we can
drop the general interpretation of $M$ in equation (\ref{eq:b_eff}), as all
properties of the galaxy are reducible to the parent halo mass (though see the
comments in the last paragraph of this section). 

Equation (\ref{eq:b_eff}) is not the end of the story, however, because it does
not tell us anything about what happens to the haloes after they have formed
and, in particular, says nothing about the timescale of any merging. In the
standard treatment  of hierarchical clustering -- the Press--Schechter (1974)
theory -- {\em all} the haloes that exist at a given stage merge immediately to
form higher mass haloes, so that in practice at each time the only haloes which
exist at all are those which just formed at that time. If one identifies the
galaxies with their hosting haloes, then automatically $z_f=z$ in the previous
formula, i.e. the galaxy merging rate is automatically assumed to be much
faster than the cosmological expansion rate. This is at the basis of what in
Paper I we called the {\em merging model}. Of course this instantaneous-merging
assumption is physically unrealistic and is related to the fact that one is
using a mass variable which is continuous, while the aggregates of matter that
form are discrete. On the other hand, it does provide a reasonable counter to
the galaxy-conserving model introduced above. 

The galaxy-conserving model (no merging) and the merging model (rapid merging)
can be regarded as two extreme pictures of how galaxy formation might proceed.
In between these two extremes, one can imagine a more general scenario in which
  galaxies neither survive forever nor merge instantaneously. The price for
this greater generality is that one requires an additional parameter to be
introduced compared to equation (\ref{eq:b_cons}). To understand how this
intermediate model is constructed, it is easiest to look at how $b(M,z)$ is
used to calculate the quantity which is really required for observational
comparisons, that is $b_{\rm eff}(z)$, which appears in equation
(\ref{eq:xifund}). Basically, one takes the `monochromatic' (i.e. single mass)
bias at each redshift (possibly with some extra parametric dependence on $z_f$
different from $z$) and then averages this monochromatic bias over the  mass
distribution of objects to obtain the `effective bias', as described in
equation (\ref{eq:b_eff}). The latter quantity is to be used to connect the
underlying mass autocorrelation function with the objects two--point function,
which also requires convolution with the redshift distribution ${\cal N}(z)$;
see equation \ref{eq:xifund}). 

As in Paper I, we can estimate the effective bias by assuming that the objects
observed in a given survey represent all haloes exceeding a certain cutoff mass
$M_{\rm min}$ at any particular redshift. In other words, we assume that there
is a selection function $\phi(z,M)=\Theta(M-M_{\rm min})$ at any $z$, where
$\Theta(\cdot)$ is the Heaviside step function. This is consistent with the
reasoning we mentioned above, that all galaxy properties are reducible to the
parent halo mass $M$. In this way, by modelling the linear bias at redshift $z$
for haloes of mass $M$ as in equation (\ref{eq:b_mono}) and by weighting it
with the theoretical mass--function $\bar n(z, M)$ which we can
self--consistently calculate using the Press--Schechter theory, we can obtain
the behaviour of $b_{\rm eff}(z)$ directly. The results for different
cosmological models are shown by the solid lines in Fig. 1, where various
choices of the minimum cutoff mass in $\bar n(z,M)$ are shown for reference. 

The behaviour of $b_{\rm eff}(z)$ can be fitted by a relation of the form 
\be
b_{\rm eff}(z)=1-1/\delta_c+[b_{\rm eff}(0)-1+1/\delta_c]/D_+(z)^\beta \;.
\label{eq:bfit}
\ee
The resulting best-fit parameters $b_{\rm eff}(0)$ and $\beta$ are reported in
Table \ref{t:fit} for different choices of initial matter fluctuation spectrum
and minimum mass. In all cases, the effective bias is a increasing function of
both redshift and $M_{\rm min}$. Notice further that a relatively strong
anti-bias $b_{\rm eff}<1$ can be produced at $z=0$ if the minimum mass is
small, because all the small haloes still existing at a late stage of the
clustering hierarchy will tend not to lie in dense regions. 

\begin{figure*} 
\centerline{\psfig{file=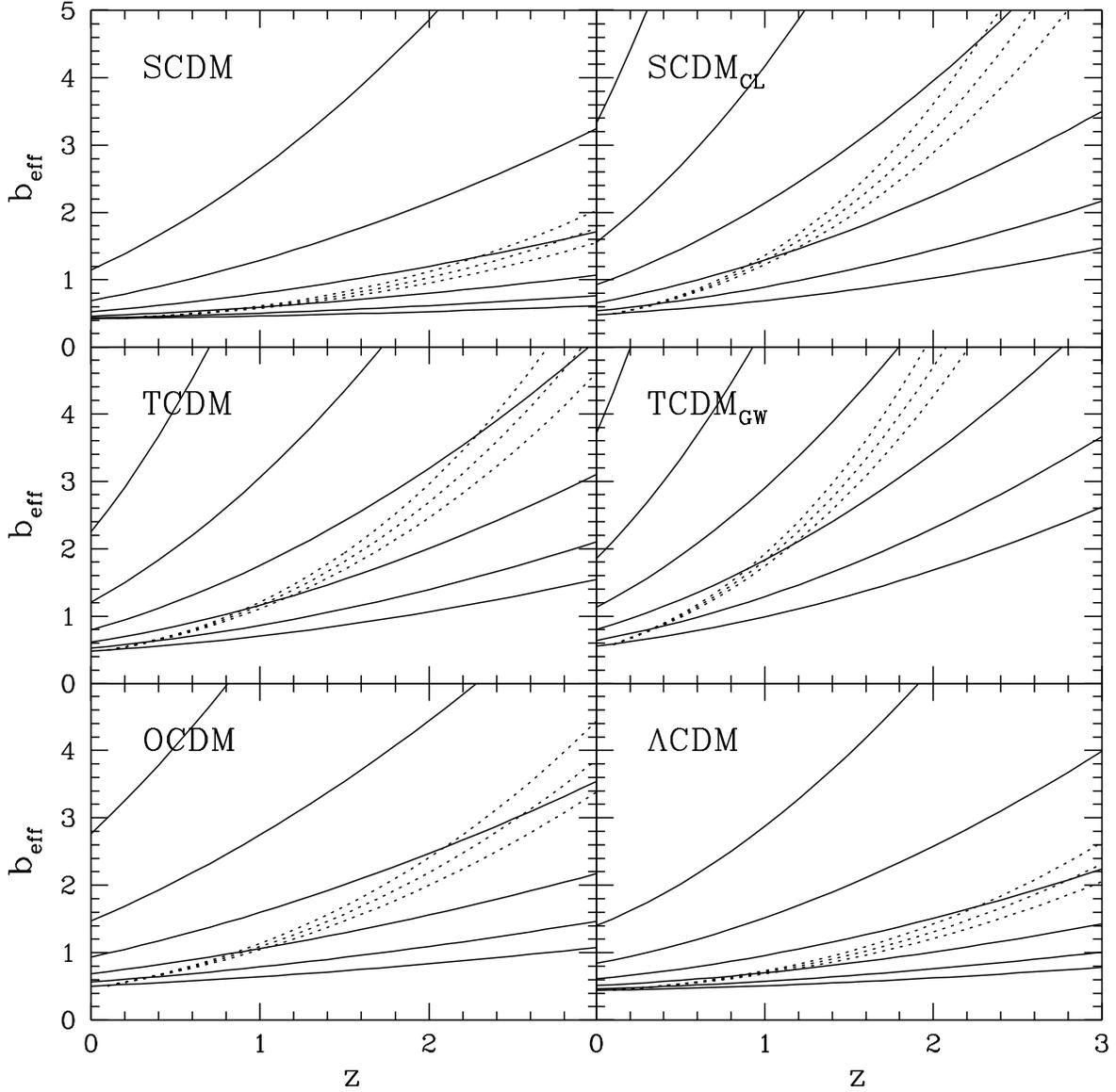,width=17.0cm,height=17.0cm}} 
\caption{
The effective bias $b_{\rm eff}$ as a function of the redshift $z$ for the
cosmological models considered in this paper. The different solid lines refer
to different values of the minimum mass $M_{\rm min}$ in the Press--Schechter
mass--function, ranging from $10^9$ to $10^{14} ~h^{-1}~M_\odot$, from bottom
to top. The dotted lines show the effects of the catalogue selection: the
results are obtained by assuming a bolometric magnitude limit $m_{_{\rm
lim}}=24$, a mass-to-luminosity ratio $M/L= 10 M_\odot/L_\odot$ and a
K+E--correction expressed by the relation ${\cal K}\log(1+z)$, with ${\cal
K}=-1,0,+1$ (from bottom to top). 
}
\end{figure*} 

\begin{table}
\centering
\caption[]{The best-fit parameters of the relation for the effective bias
(\ref{eq:bfit}) computed for different minimum mass and different
cosmological models.}
\tabcolsep 4pt
\begin{tabular}{ccccccccccccc} \\ \\ \hline \hline
 & \multicolumn{2}{c} {$SCDM$} & \multicolumn{2}{c} {$SCDM_{CL}$}
& \multicolumn{2}{c} {$TCDM$} & \multicolumn{2}{c} {$TCDM_{GW}$}
& \multicolumn{2}{c} {$OCDM$} & \multicolumn{2}{c} {$\Lambda CDM$} \\
\hline
$M_{\rm min}$ & $b_{\rm eff}(0)$ & $\beta$ & $b_{\rm eff}(0)$ & $\beta$ &
                $b_{\rm eff}(0)$ & $\beta$ & $b_{\rm eff}(0)$ & $\beta$ &
                $b_{\rm eff}(0)$ & $\beta$ & $b_{\rm eff}(0)$ & $\beta$ \\
\hline
$10^{9}\ h^{-1}M_\odot$  & 0.42 & 1.97 & 0.48 & 1.94 & 0.48 & 1.97 &
                           0.56 & 1.95 & 0.51 & 1.96 & 0.44 & 1.98 \\
$10^{10}\ h^{-1}M_\odot$ & 0.43 & 1.93 & 0.54 & 1.89 & 0.53 & 1.93 &
                           0.64 & 1.91 & 0.57 & 1.93 & 0.47 & 1.95 \\
$10^{11}\ h^{-1}M_\odot$ & 0.46 & 1.86 & 0.66 & 1.82 & 0.61 & 1.87 &
                           0.80 & 1.86 & 0.69 & 1.87 & 0.51 & 1.90 \\
$10^{12}\ h^{-1}M_\odot$ & 0.52 & 1.75 & 0.92 & 1.75 & 0.80 & 1.79 &
                           1.13 & 1.80 & 0.93 & 1.80 & 0.61 & 1.82 \\
$10^{13}\ h^{-1}M_\odot$ & 0.69 & 1.64 & 1.56 & 1.73 & 1.21 & 1.74 &
                           1.85 & 1.77 & 1.47 & 1.75 & 0.84 & 1.73 \\
$10^{14}\ h^{-1}M_\odot$ & 1.15 & 1.62 & 3.33 & 1.79 & 2.25 & 1.75 &
                           3.72 & 1.81 & 2.77 & 1.76 & 1.40 & 1.70 \\
\hline
\end{tabular}
\label{t:fit}
\end{table}

So far this argument works equally well for the calculation of $b_{\rm eff}(z)$
in the framework of the merging model. But notice that the parameters of that
model are entirely fixed because one has to match the evolution of the
clustering hierarchy to observations at the present epoch. This amounts to
matching the present-day value of $b(z)$ by relating clustering properties of
bright galaxies (e.g. the measured value of $\sigma_8$ for these galaxies) to
the analogous quantity predicted in a given model for the matter distribution.
In terms of the argument given in the preceding paragraph, this basically means
that $M_{\rm min}$, the free parameter, is fixed by requiring the present
population of galaxies to have been entirely produced by a merger-driven
hierarchy. On the other hand, we might decide that galaxies one might happen to
see at large redshifts cannot be identified with the ancestors of present-day
galaxies. In this case one can regard $M_{\rm min}$ as a free parameter, but
once a minimum mass is chosen, the value of $b_{\rm eff}$ implied at redshift
$z=0$ will probably not correspond to the value of the bias parameter measured
for any known class of objects. One must then assume that such objects are
missing in local surveys, either because they had undergone a transient
increase in their luminosity and have now faded or because they correspond to
extended objects of low surface brightness, visible at high redshift, but
invisible at small distances, due to selection effects. In Paper I we called
this model the {\em transient model} and we will adopt this nomenclature here. 

The transient model is more strongly motivated from a theoretical point of view
for QSOs rather than galaxies. Efstathiou \& Rees (1988) have used efficiency
arguments to obtain a relatively high minimum halo masses for QSOs, which may
be as large as $\sim 10^{12} {\rm M}_{\odot}$. On the other hand, Haiman \&
Loeb (1997) have suggested that short-lived quasar activity may be possible in
haloes of much lower mass than this. The transient model for quasars appears to
be consistent with observations of the clustering evolution of these objects
(e.g. La Franca, Andreani \& Cristiani 1997), but  the available QSO data do
not rule out alternative models based on the clustering of collapsed objects
which can behave in a qualitatively similar way to the transient model
(Brainerd \& Villumsen 1994; Ogawa et al. 1997; Bagla 1997b). In the galactic 
setting the choice of halo mass is much less obvious than in the case of QSOs,
and the haloes that host typical galaxies may well be much smaller than those
that host QSOs. 

We should make the point that these simple schemes do not exhaust all the
possible scenarios through which galaxies might have formed and evolved. For
example, it is quite possible that merging could play a different role at
different redshifts. Present day bright disk galaxies, for example, have
clearly not just formed at the present epoch since their properties suggest a
lack of mergers in the recent past. On the other hand, it is plausible that
galaxies at much higher redshifts, say $z\sim 2$, are undergoing merging on the
same timescale as the parent haloes. This suggests the possible applicability
of a model where rapid merging works at high redshift, but it ceases to
dominate at lower redshifts and the bias then evolves by equation
(\ref{eq:b_cons}) until now. In this context it is interesting to note that,
while $b_f$ is a free parameter in equation (\ref{eq:b_cons}), it is actually
predicted by equation (\ref{eq:bfit}), once the appropriate minimum mass is
specified. Thus matching the merging phase (\ref{eq:bfit}) onto the conserving
phase (\ref{eq:b_cons}) gives 
\be
b_f = 1 +(b_0-1) D_+^{-1}(z_f)=1-\frac{1}{\delta_c} + \left( b^*_{\rm eff}(0)
-1 + \frac{1}{\delta_c} \right) D_+(z_f)^{-\beta}
\ee
for the bias these objects would have at $z_f$ when they stopped merging. In
this equation $b^*_{\rm eff}(0)$ is to be interpreted as the effective bias the
galaxies would have now if they continued merging from $z_f$ until now; since
the galaxies do not do this, the actual present-day bias will be different.
Evolving from $z=z_f$ until $z=0$ using equation (\ref{eq:b_cons}) yields 
\be
b_0 =1 -\frac{D_+(z_f)}{\delta_c} + \left(b^*_{\rm eff}(0)-1+
\frac{1}{\delta_c}\right)
D_+^{1-\beta}(z_f)\,.
\label{eq:b0_fit}
\ee
It is interesting to speculate whether the clustering of present-day `bright'
galaxies can be explained in this picture. For example, in an Einstein-de
Sitter model, such galaxies must have $b_0\simeq 2$ if the constraint on
$\sigma_8$ from cluster abundances is correct (Eke, Cole \& Frenk 1996; Viana
\& Liddle 1996), since $\sigma_8$ measured for these objects is of order of
unit. In order to obtain $b_0\simeq 2$, we need to have the last term in
equation (\ref{eq:b0_fit}) to be of order unity. Unless $z_f$ is very large,
this means  a relatively large value of $b^*_{\rm eff}(0)$ which, from Table 2,
requires a relatively large minimum mass of order $10^{12} h^{-1} {\rm
M}_\odot$ for those models with $\sigma_8\simeq 0.5$. Such a picture would
therefore explain a large present-day value of the bias of galaxies with very
large haloes, if these galaxies could be identified with objects that for some
reason had not undergone significant merging in the recent past. Whether this
interpretation is consistent with observations of the dependence of galaxy
merger rates with redshift (e.g. Ellis 1997; Neufschaefer et al. 1997; Roche,
Eales \& Hippelein 1997) is a subject for further study. Notice that in open
models and models with non-vanishing cosmological constant the value of
$\sigma_8$ coming from the constraint on the cluster abundance is larger (see
e.g. Viana \& Liddle 1996). For example, for $\Omega_{0m}=0.4$, $\sigma_8\simeq
0.8$ is required. Consequently the present-day `bright' galaxies must have
$b_0\simeq 1.3$ and the corresponding minimum mass can be smaller. We shall not
investigate this model any further in this paper, however, as it contains
nothing that makes it qualitatively different from the previous examples. 

We can now summarize these arguments by introducing a simple unified model for
bias which incorporates all four of these previous examples as special cases.
As noticed in Paper I, all these models can be described by the equation 
\be
b_{\rm eff}(z) = b_{-1} + (b_0 - b_{-1})/ D_+(z)^\beta \;,
\label{eq:bm1}
\ee
with suitable parameters $b_0$, $b_{-1}$ and $\beta$; $b_{-1}$ may be
interpreted as the bias factor at the end of the era of cosmological expansion,
i.e. at the maximum expansion in a closed model, or as $t\rightarrow\infty$ in
an open or flat model. The particular examples we consider are: 
\begin{itemize}
\item
the {\rm unbiased} model, with $b(z)=1$;
\item
the {\rm transient} model, where the parameters are fixed by the choice of the
minimum mass (see Table \ref{t:fit}); we will use $M_{\rm min}=10^{11} h^{-1}
M_\odot$; 
\item
the {\rm merging} model, where the parameters are fixed by the value of the
bias parameter at $z=0$ ($b_0=1/\sigma_8$) and taking the bias relation for the
corresponding minimum mass; 
\item
the {\rm galaxy-conserving} model, which has $b_{-1}=\beta=1$  and
$b_0=1/\sigma_8$. 
\end{itemize}

In order to be fully consistent with our formalism, we adopt the non-linear
value of the r.m.s. fluctuation amplitude inside a sphere of $8 h^{-1}$ Mpc,
$\sigma_8^{nl}$, computed by using the PD96 method. The values of
$\sigma_8^{nl}$ for the different cosmological models are reported in the last
column of Table \ref{t:models}. 

In Fig. 2, we display the actual evolution of $b_{\rm eff}(z)$ for each of the
cosmologies and for each of these four biasing models. Notice that there is a
considerable variation in the behaviour expected depending on the cosmology
under consideration, except (of course) for the unbiased model. We shall return
to this in the next Section. 

\begin{figure*} 
\centerline{\psfig{file=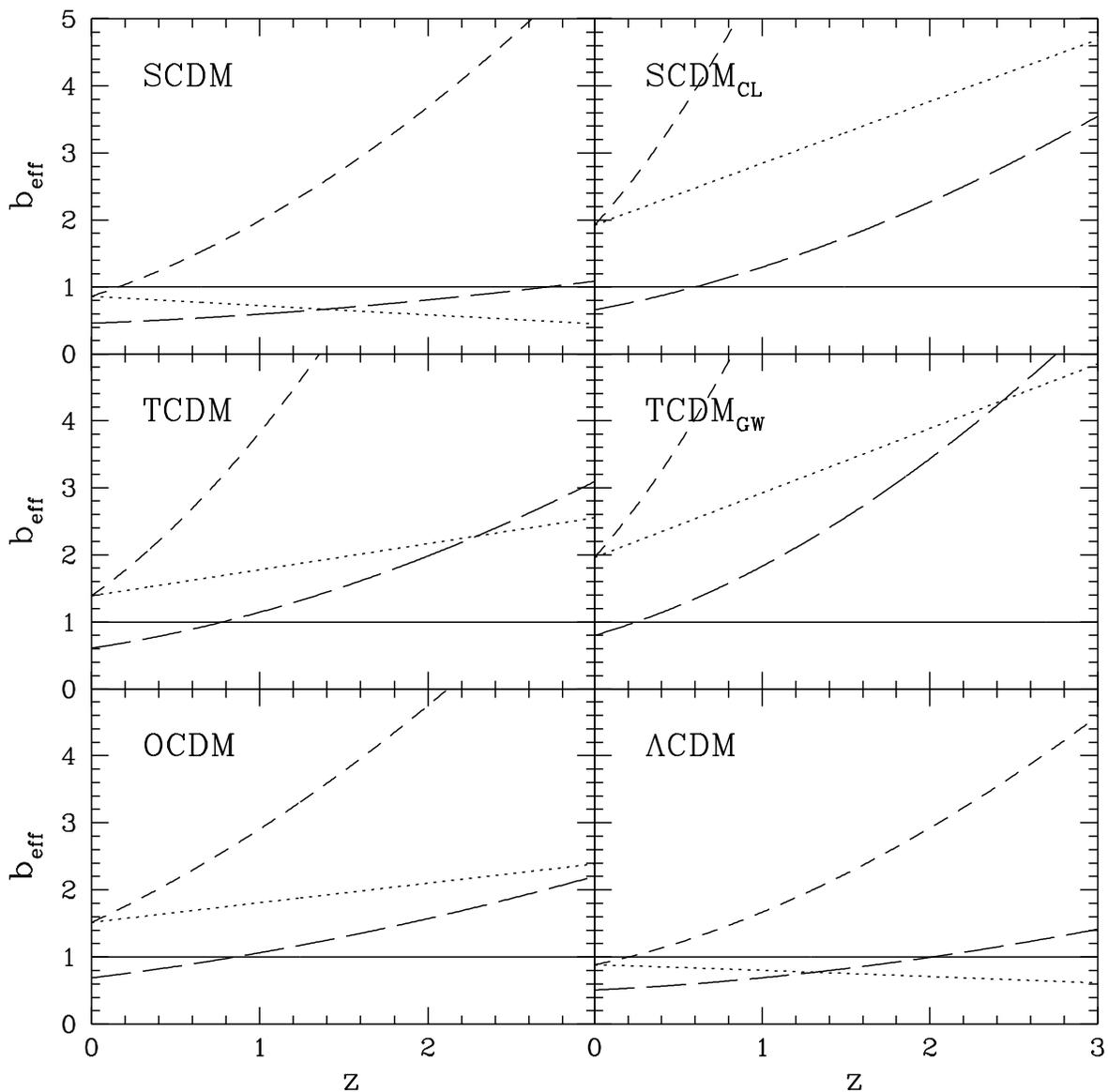,width=17.0cm,height=17.0cm}} 
\caption{
The fits to the function $b_{\rm eff}(z)$ for each cosmology discussed in this
paper and for the four different biasing models described in the text: unbiased
model (solid line); galaxy--conserving model (dotted line); merging model
(short-dashed line); transient model (long-dashed line). 
}
\end{figure*} 

As final point, we should mention how the bias factor changes when catalogue
selection effects are considered, i.e. when theoretical quantities, as the mass
$M$, are substituted with the observational ones, such as the luminosity $L$.
After choosing one of the models above, one will end up with the quantity
$b(M,z)$ to be understood as `the bias that objects of mass $M$ have at
redshift $z$'. The effective bias at the same redshift is precisely 
\be
b_{\rm eff}(z) = N(z)^{-1} \int d \ln L \,\Phi_{\rm obs}(L) b[M(L),z] \;,
\ee
where $N(z) = \int d \ln L \,\Phi_{\rm obs}(L)$ and $\Phi_{\rm obs}(L)$ is the
{\em observed}
luminosity function of the catalogue, i.e. the intrinsic luminosity function
multiplied by the catalogue selection function, which will typically involve a
cut in apparent magnitude in whatever wave-band is being used, rather than in
the somewhat idealised case we discussed above where everything corresponds to
a cut in mass $M$. Because of this cut (for magnitude or flux-limited
catalogues), in order to obtain $b_{\rm eff}$ one should need to know the
distance modulus of the galaxies at a given $z$, including all possible
K--corrections and evolutionary (E) effects in order to calculate this exactly.
Furthermore, because the bias is typically expressed as a function of mass, one
needs to know the mass-to-light ratio in the given wave-band. To test the size
of this effect we computed an illustrative example for all the cosmological
models described above, assuming a bolometric magnitude limit of $m_{_{\rm
lim}}=24$, a mass-to-luminosity ratio $M/L= 10 ~M_\odot/L_\odot$ and a
K+E--correction parametrised by the relation ${\cal K} \log (1+z)$. The results
for ${\cal K}=-1,0,+1$ are shown as dotted lines in Fig. 1. The general effect
of this `selection bias' is to exaggerate the increase of the bias factor with
redshift even further. This is particularly evident when small minimum masses
$M_{\rm min}$ are considered, though it has little effect on the quantitative
results we have obtained in the next section, and none at all on their
qualitative interpretation. We shall not therefore discuss this detail any
further in our analysis. 

\section{Simple Models of Clustering Evolution}

As we mentioned in the Introduction, theoretical interpretations of information
on clustering evolution have frequently been rather naive. In particular, many
observational results are quoted in terms of the parameter $\epsilon$ in the
following simple scaling model for the redshift evolution of the two--point
correlation function $\xi(r,z)$ at the comoving separation $r$: 
\be
\xi(r,z) = \xi(r/(1+z),0) (1 + z)^{-(3+\epsilon)}\ .
\label{eq:theor}
\ee
If the spatial dependence of the two--point function can be fitted by a
power-law with slope $\gamma$, the above relation further simplifies to 
\be
\xi(r,z) = (r/r_c)^{-\gamma} (1+z)^{-(3-\gamma+\epsilon)} \;,
\label{eq:powlaw}
\ee
where $r_c$ is a constant measuring the unit crossing of $\xi$ at $z=0$. 

Recent observational studies (e.g. Le F\`evre et al. 1996; Shepherd et al.
1997; Carlberg et al. 1997; Roche, Eales \& Hippelein 1997; Woods \& Fahlman
1997) have served to highlight the importance of understanding the validity (or
otherwise) of the simple models (\ref{eq:theor}) \& (\ref{eq:powlaw}). It is
still commonplace for observational data to be framed in terms of $\epsilon$ as
if this parameter had some unambiguous theoretical significance. For example,
Le F\`evre et al. (1996) and Shepherd et al. (1997) have reported a value of
$\epsilon \sim 1 \pm 1$ from an  analysis of galaxy clustering at moderate
redshifts. But what does this imply for theoretical models? 

Assuming that clustering grows by gravitational instability alone, the above
formulae can be interpreted in a few special cases. For $\epsilon=0$ it
reproduces the prediction of the so--called {\em stable clustering} model (cf.
Peebles 1980), while for $\epsilon=n+2=\gamma-1$, it results from the
application of linear theory in an Einstein--de Sitter universe to purely
scale--free power spectra with $P_{\rm lin}(k,0) \propto k^n$. The case where
$\epsilon=\gamma-3$ corresponds to a clustering pattern that simply expands
with the background cosmology as if the galaxies were just painted on a
homogeneous background. 

Concerning the case of stable clustering ($\epsilon=0$), one should remember
that the idea underlying the stable clustering ansatz is that, on sufficiently
small scales, gravity acts to stabilize the number of neighbours of an object
in a proper volume, after this has turned around from the universal expansion.
Numerical simulations, however, suggest that this type of dynamical regime is
only entered, if at all,  when the mass autocorrelation function is at least as
large as $\sim 100$ (e.g. Efstathiou et al. 1988; Bagla \& Padmanabhan 1996;
Padmanabhan 1996; Munshi \& Padmanabhan 1997; Jain 1997; Munshi et al. 1997),
which only occurs on scales much smaller than the dimensions of typical
surveys. Melott (1992) considered the growth of clustering in numerical
simulations for an ensemble of scale--free models. He found that the lower the
value of the spectral index $n$, the larger is the value of the parameter
$\alpha\equiv\ 3 - \gamma + \epsilon$ and that positive values of $\epsilon$
are easily allowed for in all models with $n \leq 1$. Melott's explanation for
such a fast clustering growth is as follows: stable clustering is not an upper
limit to the growth of correlations; whenever the initial conditions contain
non--vanishing large--scale power, merging makes new clusters form and their
central density increases with time, which in turn enhances the growth of
correlations. Moreover, a numerical study of the evolution of the two--point
function both for the matter and halo population has been carried out by
Col\'{\i}n, Carlberg \& Couchman (1997); they obtain a scale--dependent
$\epsilon$ parameter which is about $1$ for mass particles in an Einstein--de
Sitter universe, and lower for low--density models. A broad range of values
(ranging from $-0.2$ to $1$ in the flat case and reaching lower values in the
open case) is obtained for haloes, depending on their mean density (see also
Brainerd \& Villumsen 1994). Jain (1997) has discussed the reliability of the
general relation of equation (\ref{eq:theor}) in the context of various models.
His conclusions are that the above parametrisation for the evolution of
clustering is inaccurate in CDM--like models, for two reasons. First, because
the growth of $\xi(r,z)$ with time on intermediate scales is much faster than
the $(1+z)^{-3}$ law prescribed by stable clustering at fixed proper separation
(see also PD96) and, second, because the boundary between the linear, mildly
non-linear and stable clustering regimes, occurs at scales which rapidly
change with time. 

As one can therefore see, the theoretical interpretation of $\epsilon$ is open
to some doubt. This doubt widens when one considers the different ways
$\epsilon$ could be defined when the correlation function is not of power-law
type and the dynamical evolution is not of the self-similar form, situations
which are actually expected in realistic models. In this case one could define
$\epsilon$ to be the value at a particular proper or comoving distance $r$
where the slope $\gamma$ can be defined, perhaps the scale corresponding to
$r_0$. Alternatively, one could fit a power-law to all the data and use this to
define $\gamma$ and get $\epsilon$ that way. In general these values of
$\epsilon$ will not be equal. Likewise, $\epsilon$ at a given scale $r$ need
not be constant with time (or redshift). 

To illustrate the problem we have calculated, in Fig. 3, the behaviour of
$\epsilon$ (defined at two fixed proper separations corresponding to $1 h^{-1}$
Mpc and to the value of $r_0$ at the present epoch) for three of our models
(SCDM$_{CL}$, OCDM, $\Lambda$CDM); cf. Mo (1997). The value of $\epsilon$ at
$z$ is obtained by fitting the correlation function $\xi$ in the redshift
interval $[0,z]$. The results for the mass (solid line) are relatively constant
with redshift. Notice that for this case the value of $\epsilon$ is higher than
the stable clustering value, probably due to the effects of merging. But in any
case the scale probed here is much larger than the scale at which one expects
stable  clustering to apply in a flat Universe. In the open case, the matter
distribution actually matches $\epsilon=0$ quite closely, which is expected
because bound objects suffer no future disturbance in open models when free
expansion takes over (cf. Padmanabhan et al. 1996). The $\Lambda$CDM model is
intermediate between these two, with clustering only freezing out much later as
the expansion begins to accelerate. 

\begin{figure*} 
\centerline{\psfig{file=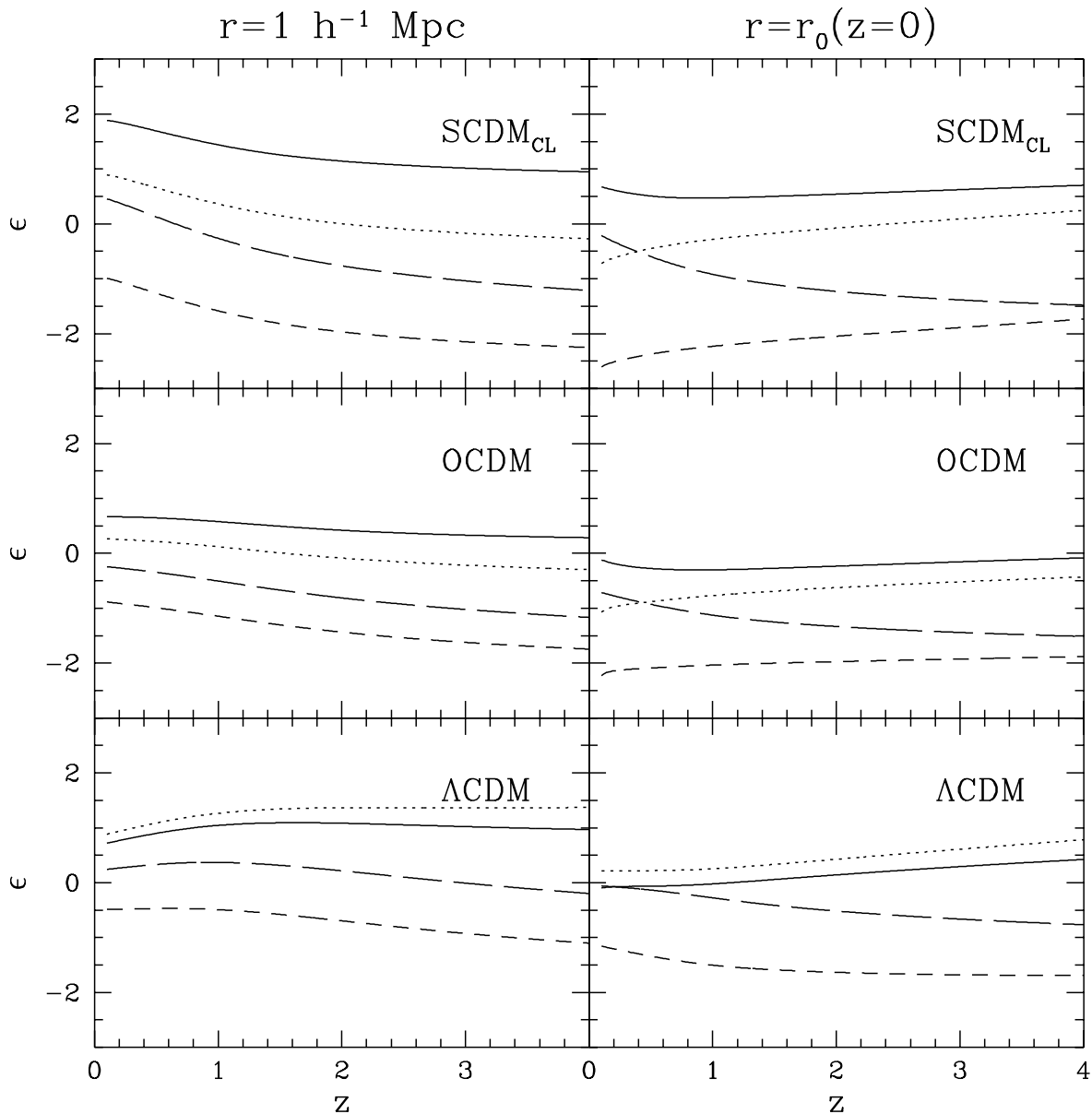,width=17.0cm,height=17.0cm}} 
\caption{
The behaviour of the scaling parameter $\epsilon$ measured at a fixed physical
length scale (equal to $1 h^{-1}$ Mpc on the left and equal to the present
correlation length $r_0$ on the right) for models with three different
background cosmologies. Different biasing models are as Fig. 2. 
}
\end{figure*} 

The situation for the {\em galaxies}, however, is much more confused than the
case for the matter. Different biasing models yield very different predicted
behaviours for $\epsilon(z)$, which suggests that the usefulness of this
parametrisation of clustering is most limited for precisely those objects which
one could actually observe. Notice also that the value of $\epsilon$ depends on
the scale at which it is measured, adding further confusion to its
interpretation. 

As well as $\epsilon$, which measures the rate of evolution, one is also
interested in what the characteristic length scale of clustering might be as a
function of redshift. One way to encode this information is via the quantity
$r_0(z)$, the distance at which the correlation function has unit amplitude.
This quantity represents a kind of characteristic scale of the clustering
pattern, so one might try to compare the sizes of individual structures with
this quantity. 

In hierarchical clustering models, the generic expectation is that this
(comoving) characteristic scale must decrease with increasing redshift. Fig. 4
demonstrates that, while this is certainly true for the distribution of mass,
it need not be true for galaxies selected with particular forms of bias. An
$r_0$ that increases with redshift is obtained in both merging and transient
models in most cases. 

\begin{figure*} 
\centerline{\psfig{file=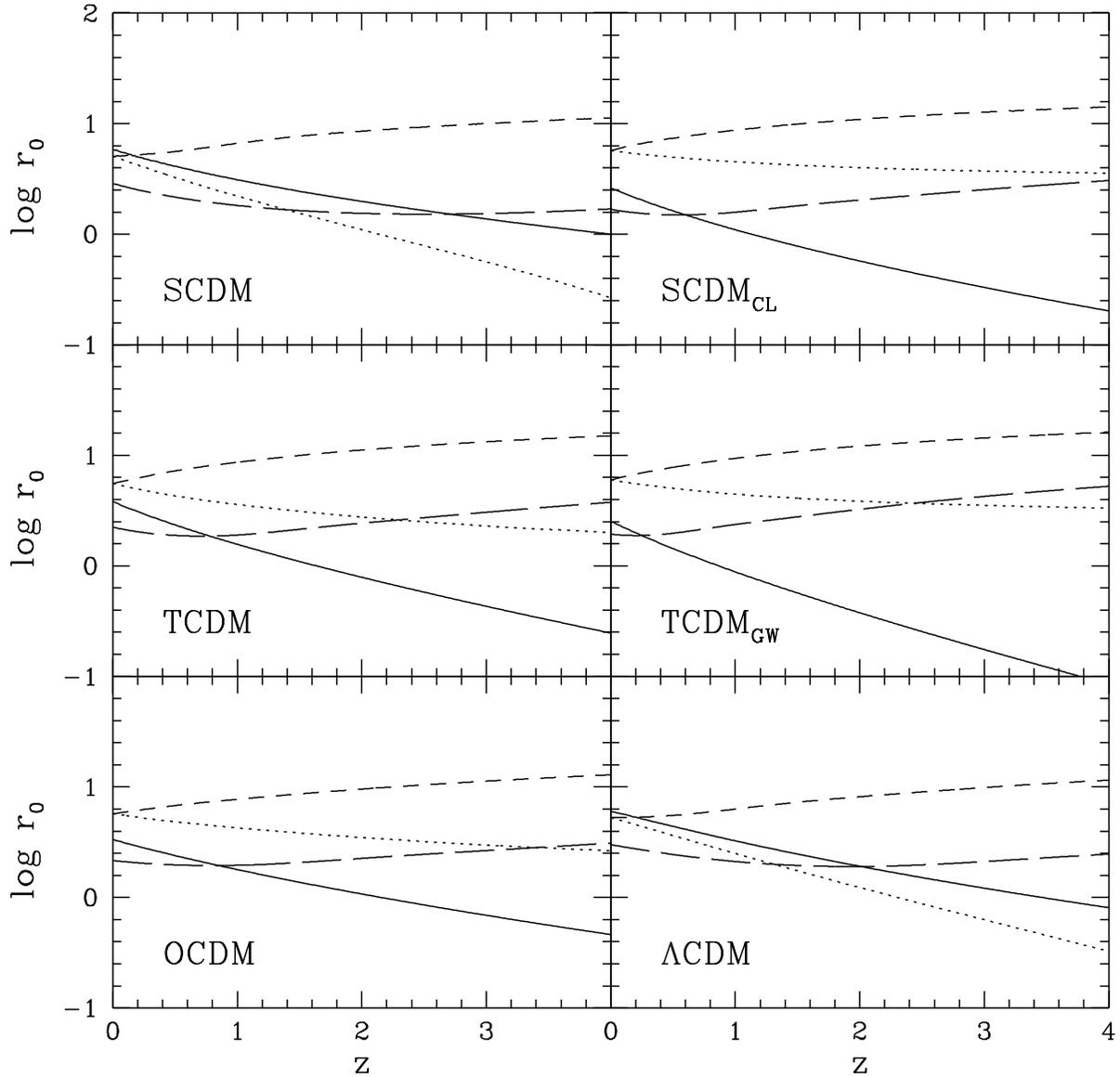,width=17.0cm,height=17.0cm}} 
\caption{
The behaviour of the (comoving) correlation length $r_0$ (in units of $h^{-1}$
Mpc) as a function of redshift for the various models considered. Different
biasing models are as in Fig. 2. Note that, although the matter correlation
length always decreases in comoving coordinates, the correlation function of
galaxies need not do so if there is significant and evolving bias. 
}
\end{figure*} 

The fundamental point that arises from these considerations concerns the
approach one adopts to test theories. In the present situation, one is
attempting to eliminate some particular well-defined models from a shortlist of
contenders. In other words, our aim is hypothesis testing. This kind of test is
best performed in the `observational plane', i.e. by computing exactly what an
observer would see in a given model universe and comparing it with what is seen
in ours. It is not useful in our view to treat the problem as one of inference,
where one tries to fit a model parameter (in this case $\epsilon$) to the
observations, particularly a parameter which is of such limited usefulness and
theoretical significance. 

\section{Applications and Results} 

\subsection{The surveys} 

In the following subsections we will apply our formalism to the correlation
analyses (both angular and projected) of three different datasets, recently
constructed for the study of the distribution of high-redshift galaxies. 

The Canada--France Redshift Survey (CFRS; Le F\`evre et al. 1996 and references
therein) consists of 591 galaxies with certain spectroscopic redshifts and
magnitudes in the range $15.5 \le I_{\rm AB} \le 22.5$. The sample covers 71
square arcminutes. The redshift distribution (Crampton et al. 1995) extends up
to $z \sim 1.6$ with more than 60 per cent of galaxies with redshift larger
then $z=0.5$. 

The Hawaii Keck K--band sample used in the study of Carlberg et al. (1997) is
an almost complete sample up to $K=20$, $I=23$ and $B=24.5$ magnitudes covering
an area of about 27 square arcminutes. The redshift distribution, presented in
Carlberg et al. (1997), is slightly changed with respect to an earlier version
of that paper, used in Paper I. Now it contains 248 galaxies, the 80 per cent
of them with redshifts between $z=0.28$ and $z=1.39$. 

The Hubble Deep Field (HDF) is a program of very deep observations made from
the Hubble Space Telescope in four passbands  and covering a field at high
galactic latitude. Different galaxy catalogues have been extracted from the HDF
(Williams et al. 1996; Clements \& Couch 1996). In their analysis, Villumsen,
Freudling \& da Costa (1997) use a total of 1732 galaxies detected in the F606W
filter which has characteristics similar to an R passband. Recently estimates
of the redshift distribution based on photometric redshifts have been obtained
by different authors (Mobasher et al. 1996; Sawicki, Lin \& Yee 1997; Connolly
et al. 1997). In order to compare our predictions with the Villumsen, Freudling
\& da Costa (1997) results, we are forced to use the same redshift distribution
adopted in that paper, i.e. 
\be
{\cal N}(z)=2.723 {z^2\over z_0^3} \exp [-(z/z_0)^{2.5}]\ ,
\label{eq:nz_hdf}
\ee
where $z_0$ is the median redshift.

The last dataset here considered is the survey for $z\sim 3$ galaxies recently
started by Steidel et al. (1996, 1998). Their observations use the $U_n$, $G$
and $R$ photometric system that is sensitive to the Lyman break in
high-redshift objects. After spectroscopic confirmation, they found 67 objects
with redshift $z>2$  in one of their fields (SSA22a+b), covering an angular
area of $8.74'\times 17.64'$. 

\subsection{The CFRS angular correlation function} 

In Fig. 5 we show the model predictions for the angular correlations of the
CFRS catalogue limited to $z<1.6$. In the same plot we also show the
observational results obtained by Hudon \& Lilly (1996) using two different
methods which likely bracket the true values: local and global determinations
are presented by open and filled squares, respectively. 

Notice that for SCDM, only the merging model is compatible with the data,
indicating that one cannot reconcile the objects in this survey with observed
galaxies at low redshift. All other biasing models fail to reproduce the data
within the SCDM framework: the predicted evolution of clustering is too strong
in this scenario. In the other cosmological models, the transient model is
always compatible with the data, and for some of them (SCDM$_{CL}$, TCDM,
TCDM$_{GW}$ and OCDM) also the unbiased model can roughly reproduce the data.
The data are incompatible with both the merging and galaxy-conserving models
for any cosmology (cf. Roukema \& Yoshii 1993). 

\begin{figure*} 
\centerline{\psfig{file=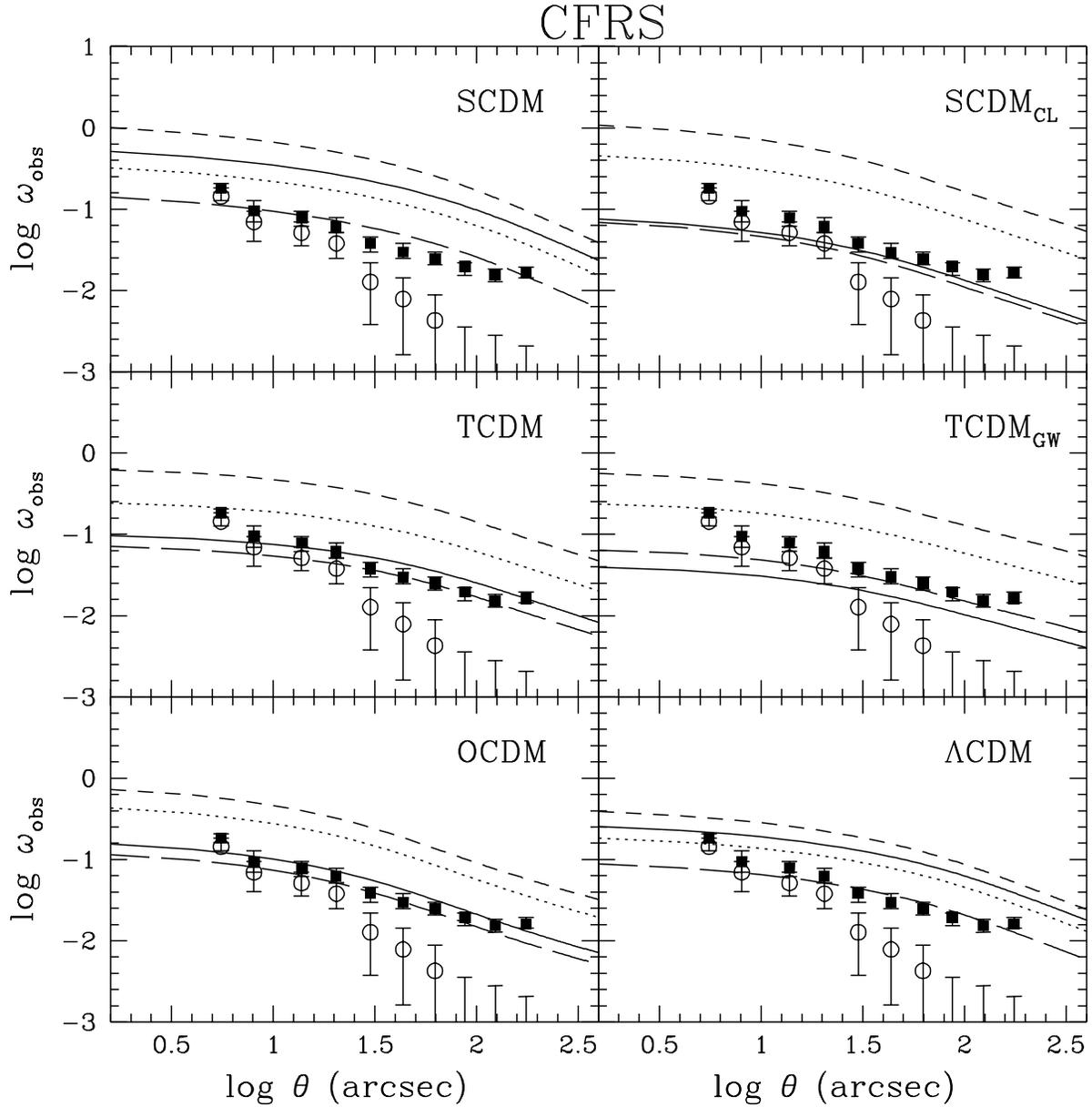,width=17.0cm,height=17.0cm}} 
\caption{
Theoretical prediction in different cosmological models for the angular galaxy
correlation function from the Canada--France Redshift Survey. The galaxies have
$z < 1.6$ and ${\cal N}(z)$ is taken from Crampton et al. (1995). Correlation
data are from Hudon \& Lilly (1996) and are obtained by using two different
methods which bracket the true values: the local and global determinations are
shown by open circles and filled squares, respectively. Different bias models
are considered, as in Fig. 2. 
}
\end{figure*} 

\subsection{Keck K-band angular correlation function}

Carlberg et al. (1997) computed the angular correlation function for the Keck
dataset limited to $z<1.6$. The comparison between these results and our
various models is shown in Fig. 6. 

The rather large errors on the observational correlations mean that
discriminatory power is less than in the previous case. Basically, all biasing
schemes are compatible with the data in any of the models, although the 
merging model predictions are uncomfortably high for both versions of SCDM and
TCDM (cf. Roukema \& Yoshii 1993). 

\begin{figure*} 
\centerline{\psfig{file=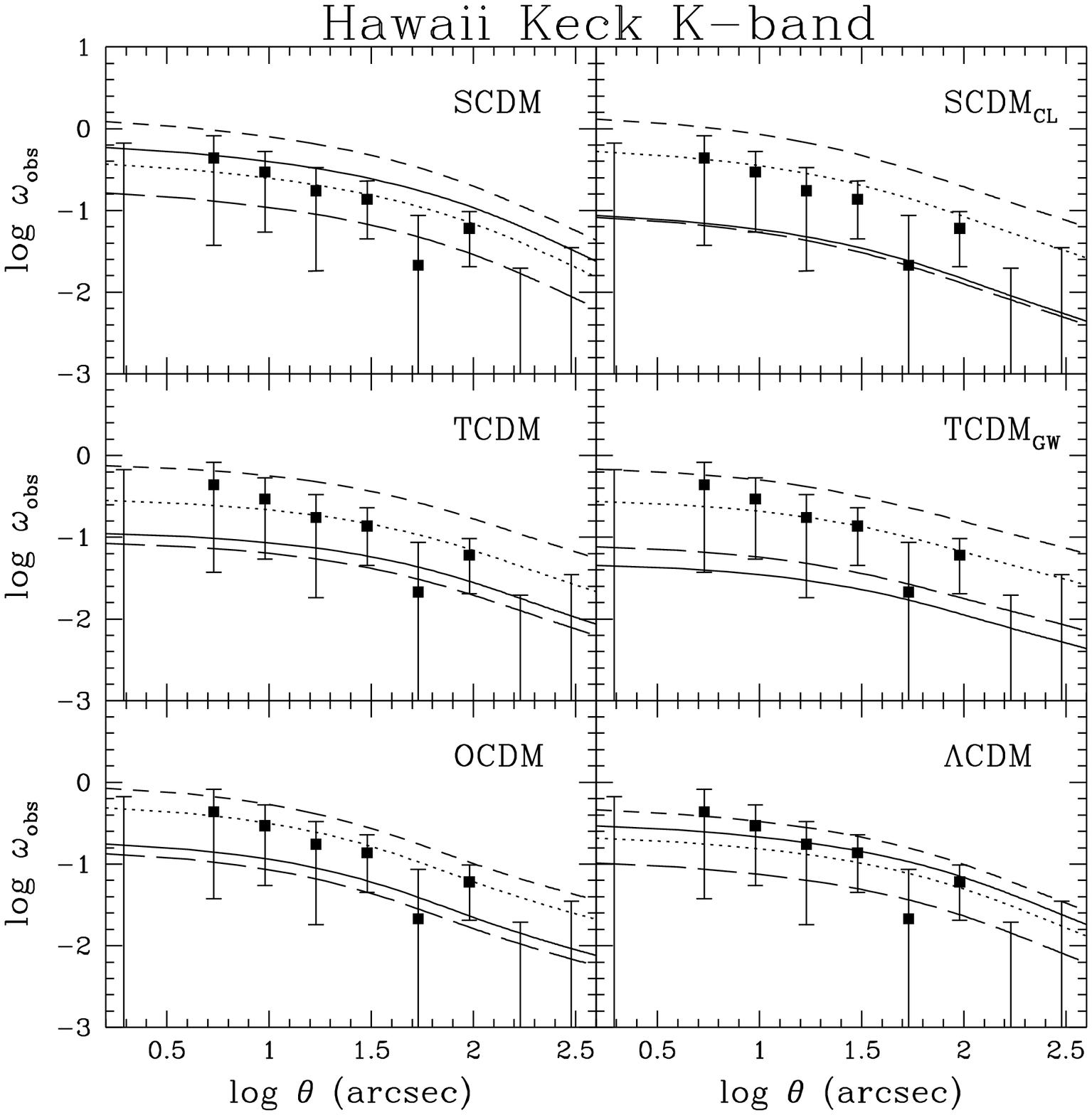,width=17.0cm,height=17.0cm}} 
\caption{
Theoretical prediction in different cosmological models for the angular galaxy
correlation function from the Hawaii K-band Survey. The galaxies are in the
redshift range $0 < z < 1.6$, with redshift distribution and correlation data
taken from Carlberg et al. (1997). The original data are corrected to take into
account the dilution produced by the uncorrelated foreground stars. Different
bias models are shown as in Fig. 2. 
}
\end{figure*} 

\subsection{Hubble Deep Field Angular Correlations}

Villumsen, Freudling \& da Costa (1997) computed the two-point angular
correlation function for the HDF survey using eight different magnitude limits,
ranging from $R=26$ to $R=29.5$. Here we prefer to compare the predictions of
our different models to the observational results only for the catalogue with
limit $R=29$, corresponding to a median redshift $z_0=1.87$ in equation
(\ref{eq:nz_hdf}). With this choice the observational results have the smaller
errorbars and are expected  to be more discriminant. In fact, the results,
shown in Fig. 7, impose impressively strong constraints on the combination of
biasing scheme and cosmological model. {\em All} combinations are excluded for
SCDM. The merging model is always excluded in any cosmology (cf. Roukema \&
Yoshii 1993). The galaxy-conserving model can fit the data only within the
$\Lambda$CDM model. The models which appear to be in best agreement with the
data are the unbiased and transient bias schemes in low density models (either
with or without a $\Lambda$-term). 

\begin{figure*} 
\centerline{\psfig{file=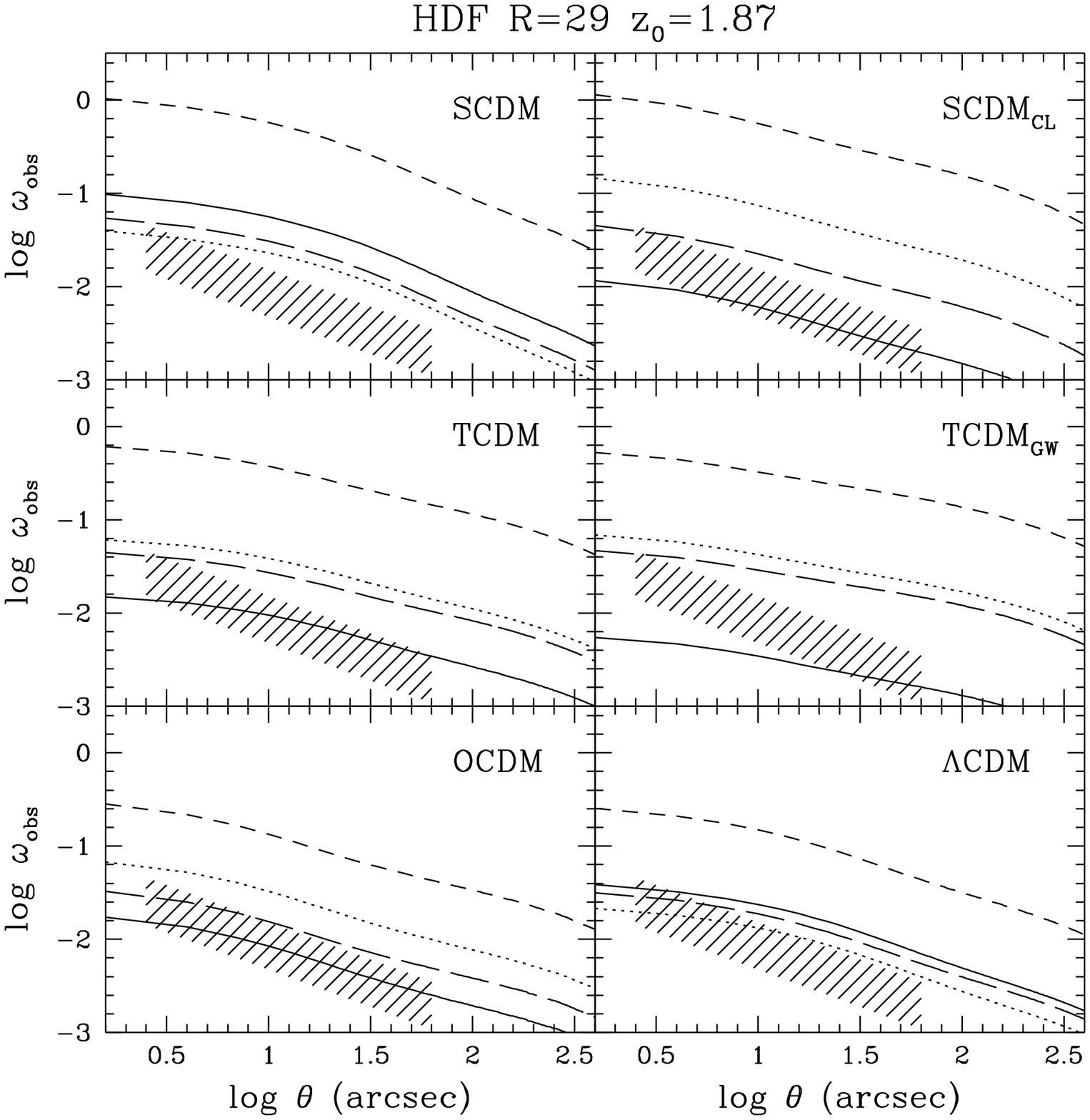,width=17.0cm,height=17.0cm}} 
\caption{
Theoretical prediction in different cosmological models for the angular galaxy
correlation function for the Hubble Deep Field. The results are for the sample
with magnitude limit $R=29$ and median redshift $z_0=1.87$. The redshift
distribution is given by equation (31). The shaded region in the plots refers
to the $1\sigma$ range allowed by the fit obtained on the observational data by
Villumsen, Freudling \& da Costa (1997). Different bias models are shown as in
Fig. 2. 
}
\end{figure*} 

\subsection{CFRS Projected Correlation Function}

The CFRS data have been analysed also in terms of the projected correlation
function by Le F\`evre et al. (1996). They divided the galaxies in three
different strips in redshift with median redshift $z\approx 0.34$, $z\approx
0.62$ and $z\approx 0.86$. We use these median redshifts to rescale in comoving
coordinates the projected separations, originally plotted in proper
coordinates. Their correlations have been computed by using $q_0=0.5$;
consequently the results have to be translated for different models because
both $w(r_p)$ and the distance $r$ depend on cosmology. For this goal we follow
Peacock (1997), particularly his discussion of the same observational dataset
in Section 4.1 of that paper and specifically using his equation (40). 

The results, presented in Fig. 8, show that the transient model is compatible
with the data for all cosmologies, though this is marginal in the case of SCDM.
The unbiased model is compatible with the data for SCDM$_{CL}$, TCDM and OCDM.
On the other hand, the merging and galaxy-conserving models are always
inconsistent. 

\begin{figure*} 
\centerline{\psfig{file=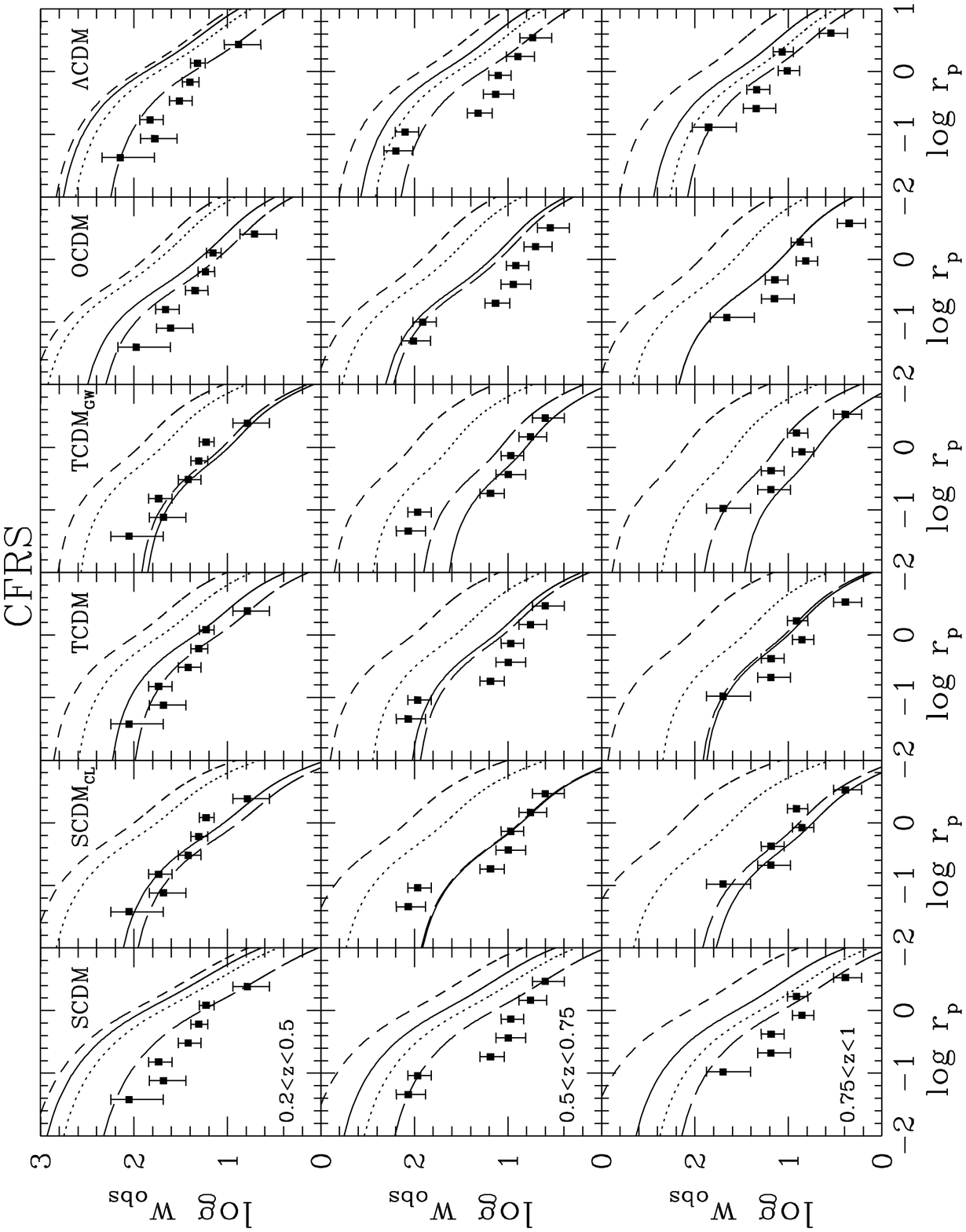,rotate=90,width=17.0cm,height=17.0cm}} 
\caption{
Theoretical prediction in different cosmological models for the projected
galaxy correlation function of the Canada--France Redshift Survey sample as a
function of the (comoving) separation $r_p$ (in units of $h^{-1}$ Mpc). The
redshift distribution is given by Crampton et al. (1995). Correlation data are
from Le F\`evre et al. (1996). Different rows refer to different strips in
redshift: $0.2 < z < 0.5$ (top),  $0.5 < z < 0.75$ (centre) and $0.75 < z < 1$
(bottom). Different bias models are shown as in Fig. 2. 
}
\end{figure*} 

\subsection{Keck K-band Projected Correlation Function}

The projected correlation function has been computed also for the Hawaii Keck
K-band survey by Carlberg et al. (1997). They present the results for four
different redshift strips, with median redshift $z\approx 0.34$, $z\approx
0.62$, $z\approx 0.97$ and $z\approx 1.39$, by adopting $q_0=0.1$. As before,
we rescale in comoving coordinates by using the median redshifts and we
translate the observational results for different cosmological models following
Peacock (1997). Notice that the observational data used here are different with
respect to those used in Paper I presented in an earlier version of the
Carlberg et al. paper. 

Interpretation of the results, reported in Fig. 9, is slightly complicated by
the strange shape of the measured correlation function at low $z$. One could
resort to a scale-dependent bias to solve this difficulty (see below), but in
any case this makes it difficult to exclude models on the basis of the results
for the low redshift bin. It is worthwhile, however, considering what one might
conclude if some of these results were subject to an unknown error. If one
accepts the points at small separations as being `accurate', then they favour
the number-conserving and merging models in all the cosmologies, and also are
consistent with an unbiased model for $\Lambda$CDM and SCDM. If instead one
discards these points and concentrates on the intermediate separation points,
they favour the transient  and unbiased models for SCDM, TCDM and OCDM. 

The data at larger redshifts have much larger errors, but it emerges robustly
that the merging model is excluded by the data for any cosmology. Generally
speaking the unbiased and transient model are reasonable fits in all cases
considered, though for SCDM the unbiased case is only marginally acceptable.
For consistency, these results at higher $z$ lead one to prefer the
interpretation that the putative problem with the low-redshift data does indeed
affect the small-separation points, rather than those at larger scale but this
argument is, of course, not rigorous. 

\begin{figure*} 
\centerline{\psfig{file=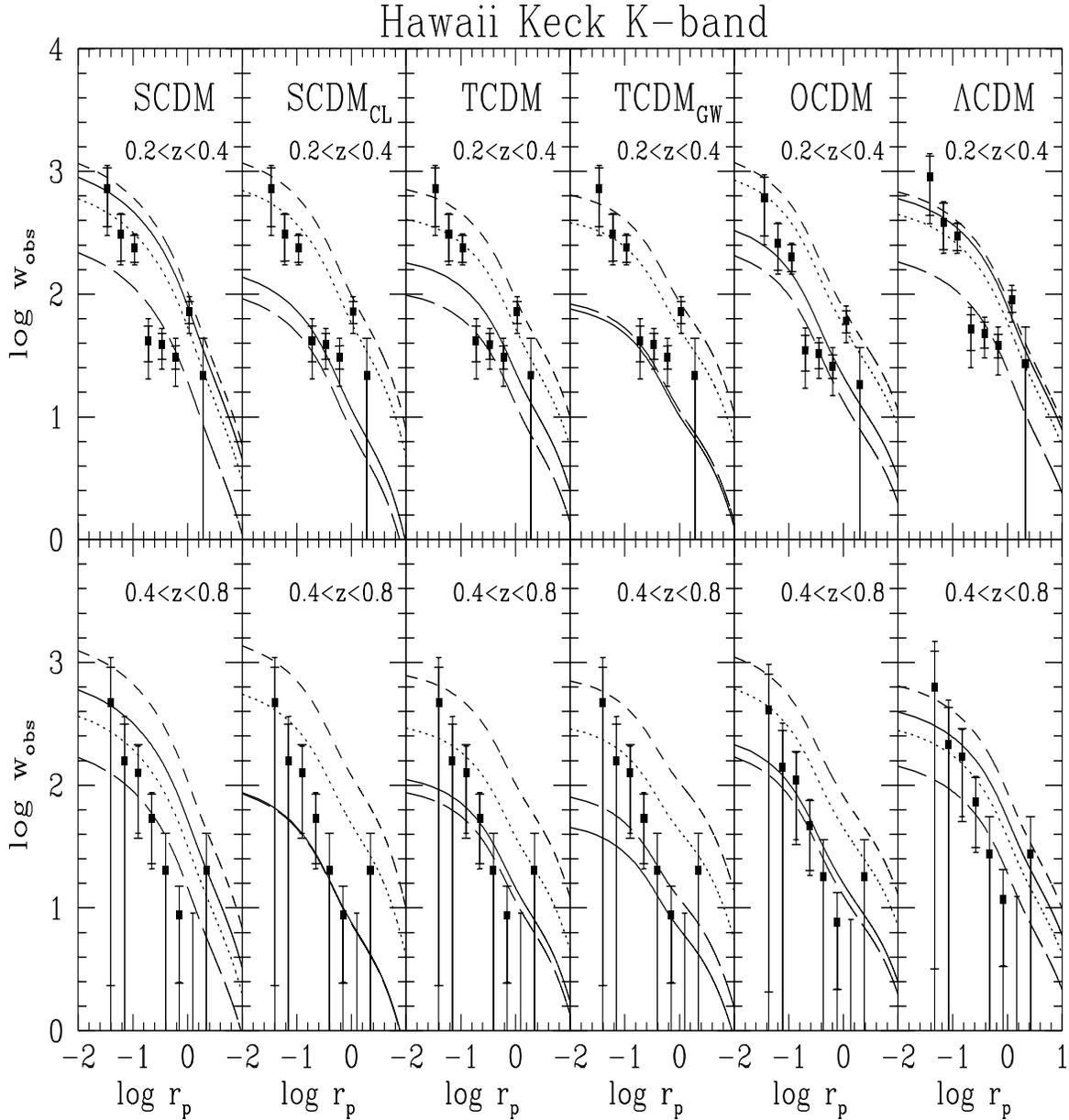,rotate=90,width=17.0cm,height=17.0cm}} 
\caption{
{\bf a.}
Theoretical prediction in different cosmological models for the projected
galaxy correlation function of the Hawaii Keck K--band survey as a function of
the (comoving) separation $r_p$ (in units of $h^{-1}$ Mpc). The redshift
distribution and correlation data are from Carlberg et al. (1997). The
1$\sigma$ bootstrap and the poissonian errorbars are shown by narrow and wide
error flags, respectively. Different rows refer to different strips in
redshift: $0.2 < z < 0.4$ (top) and $0.4 < z < 0.8$ (bottom). Different bias
models are shown as in Fig. 2. 
}
\end{figure*} 

\setcounter{figure}{8}
\begin{figure*} 
\centerline{\psfig{file=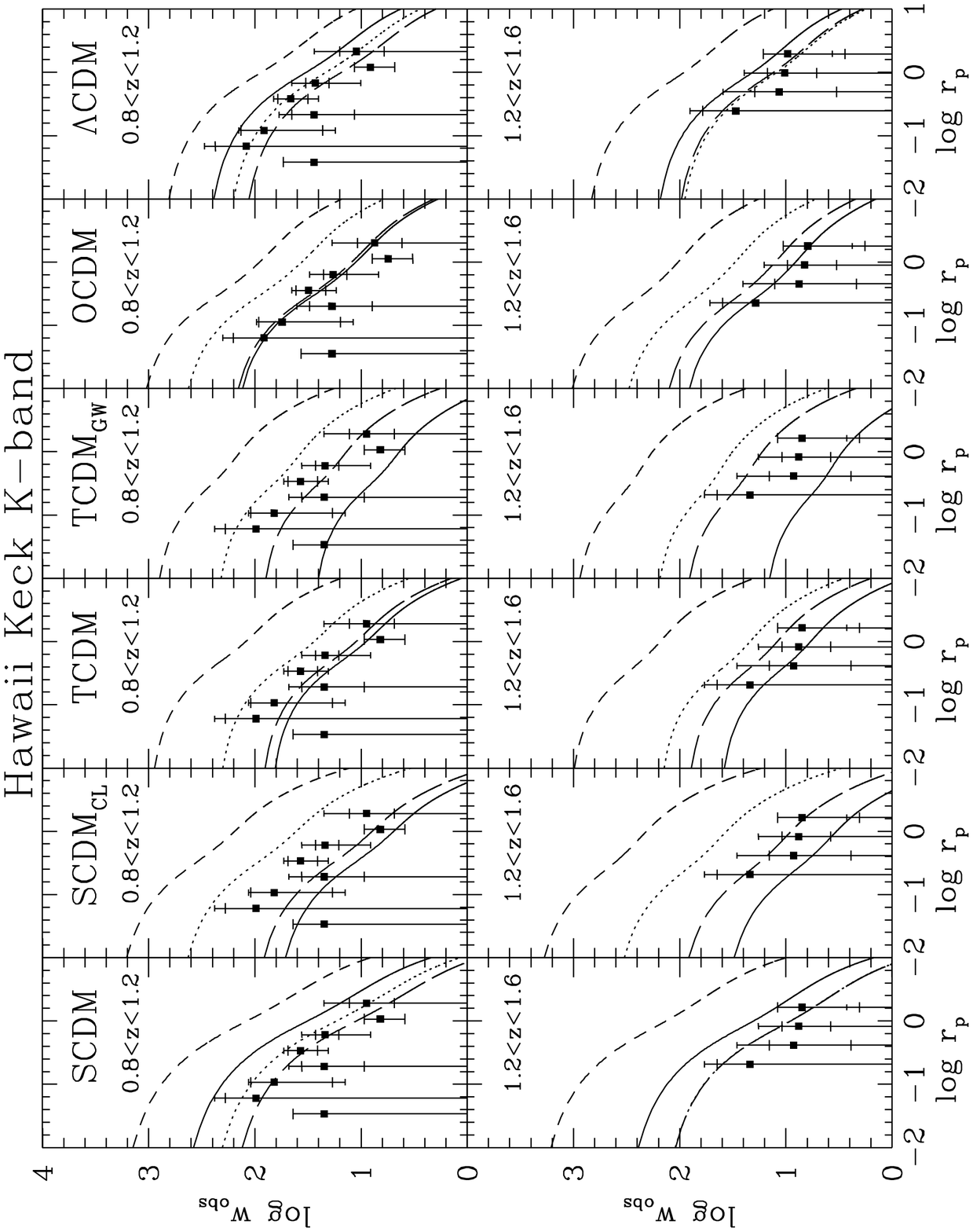,rotate=90,width=17.0cm,height=17.0cm}} 
\caption{
{\bf b.}
As Fig. 9a but for different strips in redshift: $0.8 < z < 1.2$ (top) and $1.2
< z < 1.6$ (bottom). 
}
\end{figure*} 

\subsection{Lyman-Break Galaxies}

Steidel et al. (1996, 1998) have reported evidence for the existence of a
strong concentration of galaxies at $z\sim 3$ in their angular field. This
`spike' contains 15 objects (plus one faint QSO) in a redshift bin of width
$\Delta z=0.04$. Various authors have discussed the probability of such an
object arising in particular cosmological scenarios (Mo \& Fukugita 1996; Baugh
et al. 1997; Steidel et al. 1998; Jing \& Suto 1998; Governato et al. 1998;
Bagla 1997a;  Wechsler et al. 1997; Peacock et al. 1998), reaching somewhat
equivocal conclusions. 

Although not designed for this particular problem, which can only be resolved
in an entirely satisfactory fashion using N-body simulations, the formalism we
have constructed in this paper can be used to shed qualitative light on the
concentration of Lyman-break galaxies in a very simple way. In principle, the
correct theoretical tool to this purpose would be the formula for the
probability of finding $N$ objects in a given volume, which for each model
depends on both the mean number of objects and on all the hierarchy of
correlation functions, suitably smoothed over the volume (White 1979). However,
no sound theoretical predictions exist for the evolution of the entire
hierarchy of correlation functions into the non-linear regime. We can, however,
get a useful insight by calculating the expected number of neighbours $N_R$
within a distance $R$, given the presence of an object at the origin. This is
larger than the mean number of galaxies an a randomly-selected volume by factor
of $[1+\bar{\xi}(R)]$; this factor therefore measures the average `excess'
number of galaxies that tend to accompany a given galaxy. At redshift $z$, the
quantity $N_R$ is related to the integrated mass correlation function $\bar
\xi$ by $N_R = {\overline N} [1+ b_{\rm eff}^2(z) {\bar \xi} (R,z)]$, where
$\overline N$ is the mean number of objects in a sphere of radius $R$. In order
to calculate this we need to know three different quantities: the value of
$\overline N$; the radius $R$ corresponding to the volume of the considered
bin; the appropriate model of bias. 

The value of the mean number of objects can be taken directly from the smoothed
redshift selection function, obtained by Steidel et al. (1998) from the whole
survey. From their Fig. 1, it is possible to infer that ${\overline N} \sim
4.5$ at $z \sim 3$. As for the radius $R$, the volume of the bin depends on the
cosmology because of the dependence of proper distances on angles and redshift
intervals. The translation of the angular size of the field and the width of
the redshift bin into volumes therefore depends upon the parameters $\Omega_m$
and $\Omega_\Lambda$. The bin is roughly equivalent to a sphere with $R=7.5
h^{-1}$ Mpc in a universe with $\Omega_{0m}=1$ and a factor $\sim 1.5$ larger
in the other cosmologies here considered. It is also possible that
redshift-space effects might play a role in the interpretation of these
results. In particular, it seems  quite likely (on the basis of its high
redshift) that the concentration of Lyman-break galaxies is still collapsing.
Collapse in the observer's line-of-sight would tend to enhance the
concentration observed in redshift space relative to the real space
concentration and meaning that the real space length scale of the structure
should be larger than that perceived in redshift measurements. This argues for
a larger value of $R$ than the previous values. However, Steidel et al. (1998)
and Bagla (1997a) have shown that these effects are not particularly important
on the scale of the bins chosen: the former authors put an upper limit of about
10 per cent on the redshift distortion factor. In order to allow fully for the
redshift distortions and possible background geometries, in the following
analysis we will consider, for all the models, three different values of $R$:
$R=7.5$, $R=10$ and $R=12.5 h^{-1}$ Mpc. We will call the corresponding results
$N_{7.5}$, $N_{10}$ and $N_{12.5}$, respectively. The smaller values are
probably more realistic for Einstein-de Sitter universes, whereas the larger
pair brackets the range for open and $\Lambda$-dominated models. 

The last problem is the choice of the bias model. As already discussed in
Section 3, the process of merging is expected to dominate at high redshift,
when structures are still forming hierarchically. In order to mimic the
behaviour of the Lyman-break galaxies we can therefore reasonably assume the
behaviour of  $b_{\rm eff}$ shown in Fig. 1 (with the fits reported in Table
\ref{t:fit}). The results for two different values for the minimum cutoff mass
($M_{\rm min}=10^{11}$ and $10^{12} h^{-1} M_\odot$) are reported in Table
\ref{t:steid}. 

\begin{table*}
\centering
\caption[]{The predictions of the expected number of Lyman-break galaxies
inside a sphere of radius $R=7.5$, $R=10$ and $R=12.5\ h^{-1}$ Mpc ($N_{7.5}$,
$N_{10}$ and $N_{12.5}$ respectively) for different cosmological models. Two
different values for the minimum cutoff mass ($M_{\rm min}=10^{11}$ and
$10^{12} h^{-1} M_\odot$) are used. The values of the effective bias $b_{\rm
eff}$ at redshift $z=3$ and the comoving correlation length $r_0$ 
(in units of $h^{-1}$ Mpc) are also reported.} 
\tabcolsep 4pt
\begin{tabular}{lccccccccccc} \\ \\ \hline \hline
 & \multicolumn{5}{c} {$M_{\rm min}=10^{11} h^{-1} M_\odot$} 
& & \multicolumn{5}{c} {$M_{\rm min}=10^{12} h^{-1} M_\odot$} \\
\hline
& $b_{\rm eff}(z=3)$ & $N_{7.5}$ & $N_{10}$ & $N_{12.5}$ & $r_0$ & 
& $b_{\rm eff}(z=3)$ & $N_{7.5}$ & $N_{10}$ & $N_{12.5}$ & $r_0$ \\
\hline
$SCDM$        & 1.07 & 5.2 & 4.9 & 4.8 & 1.5 && 1.72 & 6.4 & 5.6 & 5.2 & 2.7\\
$SCDM_{CL}$   & 3.50 & 6.0 & 5.4 & 5.0 & 2.5 && 6.38 & 9.5 & 7.4 & 6.3 & 4.9\\
$TCDM$        & 3.12 & 6.7 & 5.9 & 5.4 & 3.0 && 5.11 &10.5 & 8.2 & 6.9 & 5.7\\
$TCDM_{GW}$   & 5.54 & 8.0 & 6.7 & 6.0 & 4.2 && 9.30 &14.4 &10.6 & 8.6 & 7.3\\
$OCDM$        & 2.18 & 6.3 & 5.7 & 5.3 & 2.6 && 3.55 & 9.4 & 7.6 & 6.7 & 5.0\\
$\Lambda CDM$ & 1.43 & 6.0 & 5.4 & 5.1 & 2.1 && 2.24 & 8.1 & 6.8 & 6.1 & 4.1\\
\hline
\end{tabular}
\label{t:steid}
\end{table*}

It is clear from Table 3 that, for all choices of the parameter $M_{\rm min}$
and for all allowed values of $R$, the expected number is always smaller than
the observed one. However, the mean number of objects in a randomly-selected
bin at this redshift would be around 4.5. If the presence of one galaxy in the
bin is imposed then this number rises to the number given in the table. If a
typical fluctuation can raise the number from 4.5 to around 10, as it can for
models with the higher minimum mass, then a number around 15 is certainly not
an inconceivably large fluctuation. We would expect fluctuations about the mean
excess to be at least of the same order as the mean itself. Only SCDM (and,
more marginally, $\Lambda CDM$) seems to have serious problems getting close to
the value required, mainly due to the low value of the bias parameter.
Obviously, the predicted numbers decrease when larger radii are considered,
increasing the gap between model predictions and observations. Consequently,
the effect on the final results of including redshift distortions can be very
strong. 

Another comment can be made on the minimum mass. In order to have better
agreement with the Steidel et al. (1998) result, $M_{\rm min}$ has to be of
order $10^{12} h^{-1} M_\odot$ or more. This seems to indicate that the
Lyman-break galaxies should be interpretated as progenitors of massive galaxies
at the present epoch (e.g. Steidel et al. 1998) or precursor of present day
cluster galaxies (e.g. Governato et al. 1998). 

Consistency of these Lyman-break galaxy data with the model predictions also
requires that the mean number of objects estimated by Steidel et al. (1998) is
generally smaller (because of selection effects) than the mean number of these
objects predicted by the theory. This can be estimated in our formalism by
using the Press-Schechter formula to get the number of haloes more massive than
$M_{\rm min}$ at redshift $z \sim 3$. We checked that all our models satisfy
this constraint (see also Jing \& Suto 1998), but predicting precisely which
haloes give rise to a Lyman-break galaxy is beyond the scope of our theory. 

Our formalism also allows to predict the spatial correlation function $\xi_{\rm
obs}$ of the Lyman-break galaxies. To this aim we use equation
(\ref{eq:xifund}), where the redshift distribution is taken from Steidel et al.
(1998) and the bias models is chosen as before. The results obtained for a
minimum mass of $10^{12} h^{-1} M_\odot$ (that we found to be in better
agreement with the observation of the concentration of 15 galaxies at $z\sim
3$) are shown for the different cosmological models in Fig. 10. From their
N-body simulations, Governato et al. (1998) found that the effect of redshift
distortions is strong at (comoving) scales smaller than $\sim 1 h^{-1}$ Mpc
(see also Wechsler et al. 1997). For this reason we prefer to plot our results
only for larger scales. We find that the predictions for the various models are
quite different, in agreement with the analysis of Wechsler et al. (1997). The
correlation length $r_0$ (reported in Table \ref{t:steid}) ranges from $2.7
h^{-1}$ Mpc for SCDM to $7.3 h^{-1}$ Mpc for TCDM$_{GW}$. These differences,
mainly due to the large spread in the value of the bias parameter $b_{\rm eff}$
at high redshifts, seem to indicate that the measurement of the correlation
function of these objects (when reliably available) can be used to constrain
the cosmological models. 

\begin{figure*} 
\centerline{\psfig{file=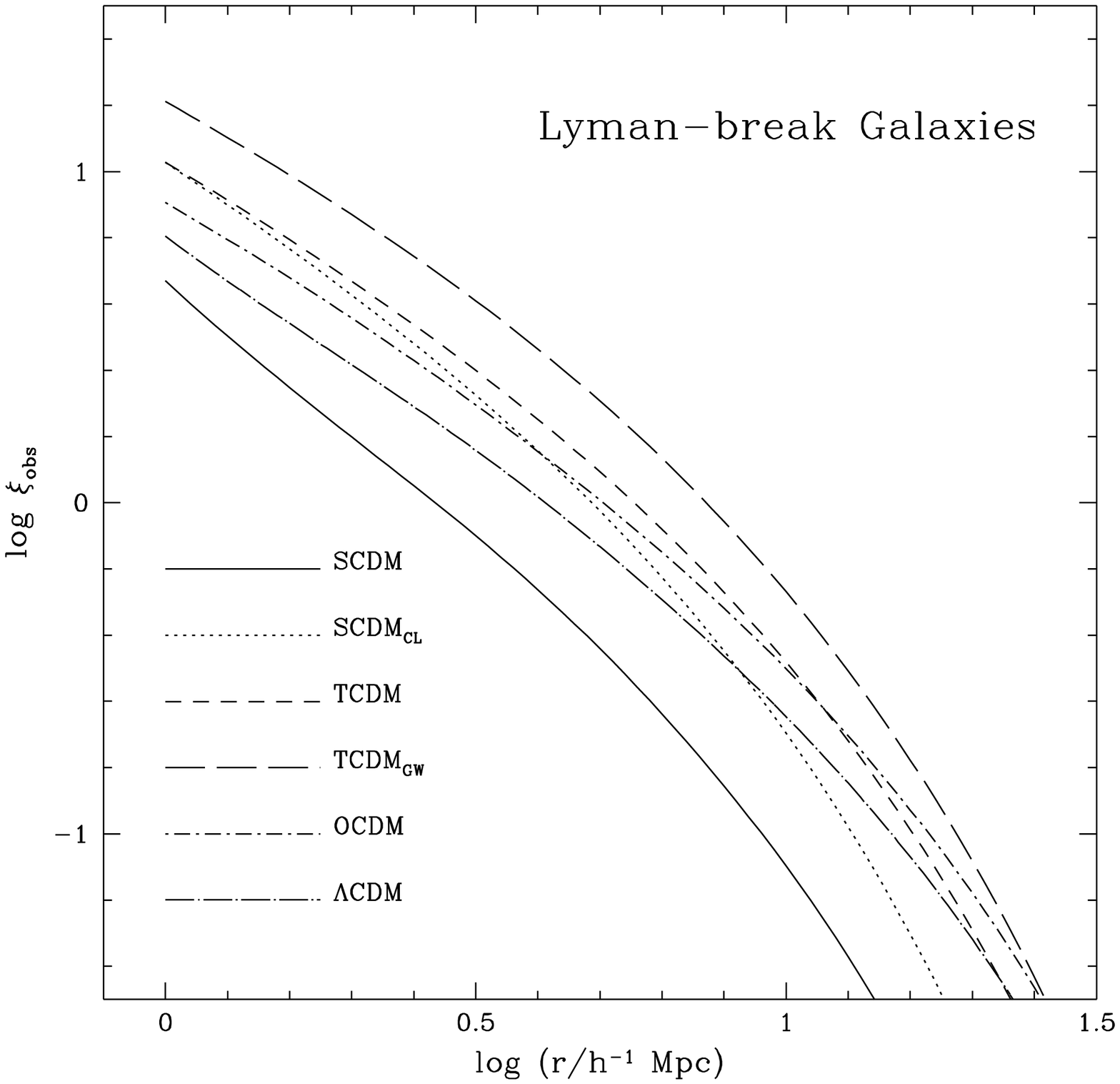,width=17.0cm,height=17.0cm}} 
\caption{
Theoretical prediction in different cosmological models for the observed
spatial correlation function of the Lyman-break galaxies as a function of the
(comoving) separation $r$ (in units of $h^{-1}$ Mpc). The redshift distribution
is taken from Steidel et al. (1998). A minimum mass of $10^{12} h^{-1} M_\odot$
is used to compute the effective bias. 
}
\end{figure*} 

\subsection{General Comments and Caveats}

The problem posed by several of these data sets concerns the {\em slope} of the
correlation function rather than its amplitude. One should not at this stage,
however, infer very negative conclusions about cosmological structure formation
scenarios on the basis of the shape. As we mentioned above, we have assumed
that the bias is modelled by a constant bias factor. As was shown by Coles
(1993), this is not the generic expectation even in local bias models and,
indeed, one expects to see a steepening of the correlation function on small
scales resulting from the introduction of non-linear terms into a generic
biasing relation of the form 
\be
\delta_{\rm n}({\bf x}; M,z) \simeq f_{M,z} [ \delta_{\rm m}({\bf x},z)]\ .
\ee
The more non-linear the function $f$, the steeper one expects the galaxy
correlation function to be compared with the matter correlations. One also
expects this phenomenon to be more prominent when the linear bias factor (which
can be thought of as the first term in a series approximation to $f$) is large,
i.e. at high $z$. However, the situation in realistic scenarios is not as clear
cut as this. In biasing models based on the properties of dark matter haloes,
we are not dealing with a generic Taylor expansion but a specific one dealing
with the relationship between haloes and mass. As shown by Catelan et al.
(1998) [see their equation (48)], the fact that the linear bias calculated
according to the Mo \& White formula is high implies that the dominant
contribution comes from halo masses much larger than $M_\star$ at the relevant
epoch. This, in turn, means that the mass field smoothed on that scale is very
close to linear. In such a case the Mo \& White result becomes more and more
accurate and higher-order corrections become negligible. So while corrections
to the linear bias formula are certainly possible, they are not necessarily
required simply because the bias is large. In any case, the observed
correlation functions do tend to be steeper than the theoretical predictions at
small separations, especially in Fig. 9a, so this might well be connected with
these effects and should not necessarily lead one to argue that none of the
models we present is compatible with the data. A non-linear bias may also play
a role in the behaviour the Lyman--break galaxies. 

It is also quite possible for the bias to be even more complicated than this.
In particular, it may be of non-local form so that the propensity of a galaxy
to form at a particular position depends not only on the density at that point,
but on the density at surrounding points. Such a non-local bias may be induced
by astrophysical effects resulting in some kind of feedback (e.g. Babul \&
White 1991; Bower et al. 1993). A non-local bias is also induced purely
dynamically, because haloes remember the conditions at their Lagrangian
birthplace (Catelan et al. 1998). 

It is also worth mentioning the somewhat surprising fact that the differences
in predictions of the cosmological models considered, while they are
significant, are not perhaps as large as one would naively imagine.  In
particular, one might have expected the $\Lambda$CDM model and OCDM model to
display the biggest differences because the  linear growth law is so different
in these cases. One can see, however, that when non-linear and bias evolution
are incorporated, these models make predictions for most of the observational
setups that are not drastically out of line with the other scenarios
considered. 

Finally, in this section we remind the reader that the correlation amplitudes
measured by observers in our theoretical universes would be even larger than
the quantities we have presented because of the effect of the amplification
bias introduced by gravitational lensing. Since most of the failed models are
excluded because they overpredict the strength of clustering anyway, this only
reinforces our conclusions. 

\section{Discussion and Conclusions} 

In this paper we have explored a number of issues arising from the
confrontation of observational evidence of high-redshift clustering against
observations. We have stressed the importance of constructing exact statistical
descriptions of clustering so that this confrontation can be carried out in an
objective and accurate way. The calculation of the statistical quantities
required to test particular models is not trivial because it demands the
inclusion of a number of different effects but, as we have shown, this can be
done when all relevant aspects are modelled systematically. 

As in Paper I, we have stressed the crucial importance of understanding more
clearly the relationship between galaxies and mass, and how this relationship
evolves with cosmic epoch. Even the simplest plausible models of bias introduce
large uncertainties into the clustering pattern predicted in different
theories. We also emphasise that these biasing schemes are probably
over-simplifications: the bias may well be non-local and/or scale-dependent and
may involve significantly more astrophysics than we have included in our
discussion. It is a first priority to understand much better the relationship
between galaxies and the underlying matter distribution, particularly the
relationship between galaxy properties and those of the parent haloes. Some
progress is clearly being made in this area by the application of
phenomenological models of hierarchical galaxy formation (e.g. Kauffmanm,
Nusser \& Steinmetz 1997). Ultimately, however, the way forward will probably
involve all-inclusive numerical simulations that can handle gravity,
hydrodynamics and star formation simultaneously. On the other hand, it is
reassuring that even the relatively small data sets available to us have
allowed sizeable chunks of the parameter space of these models to be
eliminated. In the meantime, we can be reasonably confident that further data
will lead to stronger constraints on the simple models available at present. 

We have compared some currently fashionable models of structure formation with
the available observational data using the statistical tools mentioned in the
previous paragraph. The present data have fairly large experimental errors, but
do offer significant power to discriminate between models. This situation can
only improve as more and better high-redshift data are accumulated. In
particular we found Hubble Deep Field (HDF) data to be highly discriminatory.
More data of this type, such as is anticipated from the proposed further deep
surveys with HST, would be extremely useful. 

The details of the comparison of observations with data are discussed in the
previous Section, but it is worth emphasizing a few general inferences that can
be drawn by taking all the results together. First, we can conclude that if the
`correct' model of large-scale structure is indeed one of those we have
discussed here, then both the rapid merging and galaxy-conserving models of
galaxy formation are excluded by the data. The second point is that those
cosmologies that can reproduce the observed abundance of rich clusters can also
match the galaxy clustering observations, but only if galaxies are no more than
moderately biased at redshifts of order unity. Since the models we are testing
involved a complicated interplay of the various components (background
cosmology, perturbation spectrum, biasing scheme, etc.) it is difficult to draw
deeper conclusions from the data about any one of these components. In
particular, one might have hoped that the rate of evolution of galaxy
clustering might lead one more-or-less directly to the value of the density
parameter, $\Omega$. Although the available data show no strong preference for
either high or low values of $\Omega$, the OCDM model does seem to fit both
amplitude and shape of the available marginally better than models with a
higher density; this is particularly so at relatively low redshifts. On the
other hand, the data do generally prefer a value of the bias parameter of order
unity. A low value for the bias parameter of bright galaxies tends, on other
grounds, to favour $\Omega<1$ (e.g. Peacock 1997; Coles \& Ellis 1997). 

Finally, we stress that constraints emerging from clustering arguments, like
those we have presented here, are significantly more robust than those based
solely on number-densities, which are very sensitively dependent on assumptions
about the halo parameters and galaxy formation efficiency. Quantities based on
overdensities, such as $\xi(r)$, are constructed to be independent of the
underlying number-density of objects and one can, at least in principle, use
them to make reliable predictions even when the predicted number-density of
objects is uncertain. For this reason, we expect many useful constraints on
models to derive from ongoing and planned observational surveys. 

\section*{Acknowledgments.}
R.G. Carlberg and O. Le F\`evre are warmly thanked for providing us with
electronic versions of their results. We are grateful to S. Arnouts for helpful
discussions, to Y. Jing for useful comments about the Lyman-break galaxies and
to John Peacock for clarifying for us the interpretation of the PD formalism at
$z\neq 0$. We also thank Bepi Tormen and Fabio Governato for general comments.
LM thanks the Astronomy Unit at QMWC for their hospitability. PC is grateful to
the Dipartimento di Astronomia at the Universit\`{a} di Padova for hospitality
during a visit. Italian MURST is acknowledged for partial financial support. PC
is a PPARC Advanced Research Fellow.


\begin{thebibliography}{}
\bibitem[]{} Babul A., White S.D.M., 1991, MNRAS, 253, 31
\bibitem[]{} Bagla J.S., 1997a, MNRAS, in press, astro-ph/9709230
\bibitem[]{} Bagla J.S., 1997b, preprint, astro-ph/9711081
\bibitem[]{} Bagla J.S., Padmanabhan T., 1996, ApJ, 469, 470
\bibitem[]{} Bardeen J.M., Bond J.R., Kaiser N., Szalay A.S., 1986, ApJ, 304,
15
\bibitem[]{} Baugh C.M., Cole S., Frenk C.S., Lacey C.G., 1997, preprint,
astro-ph/9703111
\bibitem[]{} Bower R.C., Coles P., Frenk C.S., White S.D.M., 1993, ApJ, 405,
403
\bibitem[]{} Brainerd T.G., Villumsen J.V., 1994, ApJ, 431, 477
\bibitem[]{} Bunn E.F., White M., 1997, ApJ, 480, 6
\bibitem[]{} Carlberg R.G., Cowie L.L., Songaila A., Hu E.M., 1997, ApJ, 484,
538
\bibitem[]{} Carroll S.M., Press W.H., Turner E.L., 1992, ARA\&A, 30, 499
\bibitem[]{} Catelan P., Lucchin F., Matarrese S., Porciani C., 1998, MNRAS,
in press, astro-ph/9708067
\bibitem[]{} Clements D.L., Couch W.J., 1996, MNRAS, 280, L43
\bibitem[]{} Coles P., 1993, MNRAS, 262, 1065.
\bibitem[]{} Coles P., 1996, Contemp. Phys., 37, 429
\bibitem[]{} Coles P., Ellis G.F.R., 1997, Is the Universe Open or Closed?
Cambridge University Press, Cambridge
\bibitem[]{} Col\'{\i}n P., Carlberg R.G., Couchman H.M.P., 1997, ApJ, 490, 1
\bibitem[]{} Connolly A.J., Szalay A.S., Dickinson M., SubbaRao M.U., Brunner
R.J., 1997, ApJ, 486, 11
\bibitem[]{} Crampton D., Le F\`evre O., Lilly S.J., Hammer F., 1995, ApJ,
455, 96
\bibitem[]{} Dekel A., 1986, ComAp, 11, 235
\bibitem[]{} Dekel A., Rees M.J., 1987, Nat, 326, 455
\bibitem[]{} Efstathiou G., Frenk C.S., White S.D.M., Davis M., 1988, MNRAS,
235, 715
\bibitem[]{} Efstathiou G., Rees M.J., 1988, MNRAS, 230, 5p
\bibitem[]{} Eke V.R., Cole S., Frenk C.S., 1996, MNRAS, 282, 263
\bibitem[]{} Ellis R.S., 1997, ARA\&A, 35, 389
\bibitem[]{} Fry J.N., 1996, ApJ, 461, L65
\bibitem[]{} Gheller C., Pantano O., Moscardini L., 1998, MNRAS, in press,
astro-ph/9710096
\bibitem[]{} Governato F., Baugh C.M., Frenk C.S., Cole S., Lacey C.G., Quinn
T., Stadel J., 1998, in press
\bibitem[]{} Haiman Z., Loeb A., 1997, ApJ, submitted, astro-ph/9710208
\bibitem[]{} Hamilton A.J.S., Kumar P., Lu E., Mathews A., 1991, ApJ, 374, L1
\bibitem[]{} Hudon J.D., Lilly S.J., 1996, ApJ, 469, 519
\bibitem[]{} Jain B., 1997, MNRAS, 287, 687
\bibitem[]{} Jain B., Mo H.J., White S.D.M., 1995, MNRAS, 276, L25
\bibitem[]{} Jenkins A. et al. 1997, ApJ, submitted, astro-ph/9709010
\bibitem[]{} Jing Y.P., Suto Y., 1998, ApJ, 494, L5
\bibitem[]{} Kauffmann G., Nusser A., Steinmetz M., 1997, MNRAS, 286, 795
\bibitem[]{} La Franca F., Andreani P., Cristiani S., 1997, preprint,
astro-ph/9711048
\bibitem[]{} Le F\`evre O., Hudon D., Lilly S.J., Crampton D., Hammer F.,
Tresse L., 1996, ApJ, 461, 534
\bibitem[]{} Lidsey J.E., Coles P., 1992, MNRAS, 258, L57
\bibitem[]{} Lilje P.B., 1992, ApJ, 386, L33
\bibitem[]{} Lucchin F., Matarrese S., 1985, Phys. Rev., D32, 1316
\bibitem[]{} Lucchin F., Matarrese S., Mollerach S., 1992, ApJ, 401, L49
\bibitem[]{} Matarrese S., Coles P., Lucchin F., Moscardini L., 1997, MNRAS,
286, 115 (Paper I)
\bibitem[]{} Melott A.L., 1992, ApJ, 393, L45
\bibitem[]{} Mo H.J., 1997, to appear in Proceedings of Ringberg Workshop
on Large-scale Structure, ed. D. Hamilton (Kluwer, Dordrecht),
astro-ph/9702218
\bibitem[]{} Mo H.J., Fukugita M., 1996, ApJ, 467, L9
\bibitem[]{} Mo H.J., White S.D.M., 1996, MNRAS, 282, 347
\bibitem[]{} Mobasher B., Rowan-Robinson M., Georgakakis A., Eaton N., 1996,
MNRAS, 282, L7
\bibitem[]{} Moessner R., Jain B., Villumsen J.V., 1998, MNRAS, 294, 245
\bibitem[]{} Munshi D., Chiang L.Y., Coles P., Melott A.L., 1997, MNRAS, in
press, astro-ph/9707259
\bibitem[]{} Munshi D., Padmanabhan T., 1997, MNRAS, 290, 193
\bibitem[]{} Neuschaefer L.W., Im M., Ratnatunga K.U., Griffiths R.E.,
Casertano S., 1997, ApJ, 480, 59
\bibitem[]{} Nusser A., Davis M., 1994, ApJ, 421, L1
\bibitem[]{} Ogawa T., Roukema B.F., Yamashita K., 1997, ApJ, 484, 53.
\bibitem[]{} Padmanabhan T., 1996, MNRAS, 278, L29
\bibitem[]{} Padmanabhan T., Cen R., Ostriker J.P., Summers F.J., 1996, ApJ,
466, 604
\bibitem[]{} Peacock J.A., 1997, MNRAS, 284, 885
\bibitem[]{} Peacock J.A., Dodds S.J., 1994, MNRAS, 267, 1020
\bibitem[]{} Peacock J.A., Dodds S.J., 1996, MNRAS, 280, L19 (PD96)
\bibitem[]{} Peacock J.A., Jimenez R., Dunlop J.S., Waddington I.,
Spinrad H., Stern D., Dey A., Windhorst R.A., 1998, preprint,
astro-ph/9801184
\bibitem[]{} Peebles P.J.E., 1980, The large--scale structure of the Universe.
Princeton University Press, Princeton
\bibitem[]{} Porciani C., 1997, MNRAS, 290, 639
\bibitem[]{} Press W.H., Schechter P., 1974, ApJ, 187, 425
\bibitem[]{} Roche N., Eales S., Hippelein H., 1997, preprint, astro-ph/9711068
\bibitem[]{} Roukema B.F., Yoshii Y., 1993, ApJ, 418, L1
\bibitem[]{} Roukema B.F., Peterson B.A., Quinn P.J., Rocca-Volmerange B.,
1997, MNRAS, 292, 835
\bibitem[]{} Sawicki M.J., Lin H., Yee H.K.C., 1997, AJ, 113, 1
\bibitem[]{} Shepherd C.W., Carlberg R.G., Yee H.K.C., Ellingson E., 1997, ApJ,
479, 82
\bibitem[]{} Steidel C.C., Giavalisco M., Pettini M., Dickinson M., Adelberger
K.L., 1996, ApJ, 462, L17
\bibitem[]{} Steidel C.C., Adelberger K.L., Dickinson M., Giavalisco M.,
Pettini M., Kellogg M., 1998, ApJ, 492, 428
\bibitem[]{} Sugiyama N., 1995, ApJS, 100, 281
\bibitem[]{} Viana P.T.P., Liddle A.R., 1996, MNRAS, 281, 323
\bibitem[]{} Villumsen J.V., 1996, MNRAS, 281, 369
\bibitem[]{} Villumsen J.V., Freudling W., da Costa L.N., 1997, ApJ, 481, 578
\bibitem[]{} Vittorio N., Matarrese S., Lucchin F., 1988, ApJ, 328, 69
\bibitem[]{} Wechsler R.H., Gross M.A.K., Primack J.R., Blumenthal G.R., Dekel
A., 1997, preprint, astro-ph/9712141 
\bibitem[]{} White M., Viana P.T.P., Liddle A.R., Scott D., 1996, MNRAS, 283,
107
\bibitem[]{} White S.D.M., 1979, MNRAS, 186, 145
\bibitem[]{} Williams R.E. et al., 1996, AJ, 112, 1335
\bibitem[]{} Woods D., Fahlman G.G., 1997, ApJ, 490, 11
\bibitem[]{} Zel'dovich Ya.B., 1970, A\&A, 5, 84
\end{thebibliography}
\end{document}